\documentclass[twocolumn]{jpsj3}

\title{Theory of Impurity Effects on the Spin Nematic State}

\author{Junji TAKANO\thanks{E-mail address: takano@issp.u-tokyo.ac.jp}  and Hirokazu TSUNETSUGU} 

\inst{Institute for Solid State Physics, University of Tokyo, 
Kashiwanoha 5-1-5, Kashiwa, Chiba 277-8581}
\abst{
The effect of magnetic bond disorder in otherwise antiferro nematic 
ordered system is investigated. 
We introduced triangular-shaped ferromagnetic bond disorder in the S=1 
bilinear-biquadratic model on a triangular lattice.
It is shown that the coupling between the impurity magnetic moment 
and nonmagnetic excitation in the bulk yields single-moment anisotropy 
 and long-range anisotropic interaction between impurity magnetic 
 moments. This interaction can induce unconventional 
 spin-freezing phenomena observed in triangular magnet, ${\rm NiGa_2S_4}$.}
 
\kword{antiferromagnets, triangular lattice, spin nematics, quadrupolar ordering, 
 impurity effects, spin glass}

\usepackage{braket}
\usepackage{amsmath}
\usepackage{amssymb}
\usepackage{stmaryrd}
\usepackage{bm}
\usepackage{multirow}

\allowdisplaybreaks

\makeatletter
\@addtoreset{equation}{section}
\makeatother

\newcommand{\figref}[1]{Fig. \ref{#1}}
\newcommand{\tabref}[1]{Table \ref{#1}}
\newcommand{\eref}[1]{eq. (\ref{#1})}
\newcommand{\erefn}[1]{(\ref{#1})}

\begin{document}
\maketitle
\section{Introduction}

Diverse novel low-energy behaviors of geometrically frustrated magnets have 
attracted much attention\cite{Ramirez}.
The central issue is the possibility of a spin liquid, namely a quantum 
disordered state 
where magnetic long-range order is destroyed by frustration and quantum 
fluctuation\cite{Diep}.
This idea was first proposed by Anderson for a 
Heisenberg antiferromagnet on the triangular lattice 
 \cite{Anderson}. 
Although subsequent numerical works showed the presence of magnetic long 
range order for that model \cite{Triangle},
 the possibility of spin liquid states have been 
intensively studied, both experimentally and theoretically. 
Recently various compounds such as organic 
$\kappa\mathchar`-{\rm (BEDT\mathchar`-TTF)_2Cu_2(CN)_3}$\cite{Organic} and ${\rm NiGa_2S_4}$
\cite{Science} as well as ${\rm SrC_{9-p}Ga_{12-9p}O_{19}}$\cite{Kagome} are 
found to exhibit spin-liquid-like behaviors. 
In addition to their "spin liquid" states, spin glass states have been widely observed in 
geometrically frustrated magnets\cite{Ramirez}, and this may be induced by 
a small amount of quenched disorder. 
These spin glass states could be ascribed to coexistence of intrinsic geometrical frustration 
and extrinsic frustration induced by disorder, 
but are not well understood theoretically. 
Therefore this is a challenging problem.
Further it is interesting that these spin glass states are often 
observed to accompany spin liquid behavior. This implies that these two effects 
are closely related. One typical example of this coexistence is the case of the triangular 
lattice antiferromagnet ${\rm NiGa_2S_4}$ and this is the issue of this paper.

In the layered chalcogenide ${\rm NiGa_2S_4}$, magnetic Ni ions form a perfect 
triangular lattice well separated by GaS polyhedra, and therefore the Ni 
magnetism has a quasi-two-dimensional nature\cite{Science}. Each Ni ion has 
the electronic configuration of $t_{2g}^6e_g^2$ and has spin $S=1$ 
formally with no anisotropy, which is consistent with the nearly 
isotropic susceptibility. 
Several low-temperature properties indicate that this system is a good 
candidate of gapless spin liquid. 
First, neutron scattering experiments revealed only short-range correlations of 
Ni spins even below $\sim$20K. The correlation length 
saturates to $\xi \sim 20\mathrm{\AA}$, 
corresponding to seven times the inplane lattice constant\cite{Science}
and very short. Second, 
magnetic specific heat shows a power-law dependence $C_M\propto T^2$ in the 
temperature regime 
$0.35\mathchar`-4$K, which signals gapless and linearly dispersive modes of excitations\cite{Science}. 
Lastly, magnetic susceptibility approaches a finite value as temperature approaches
 0K, indicating the absence of a finite spin gap\cite{Science}.

To clarify the origin of this gapless "spin liquid" behavior in 
${\rm NiGa_2S_4}$, Tsunetsugu and Arikawa proposed a scenario of 
antiferro nematic order. This is equivalent to an antiferro spin quadrupolar 
(AFQ) order, where order parameters are quadrupole moments, 
 $Q_{\mu\mu'}=\frac{1}{2}\langle S^{\mu}S^{\mu'}+S^{\mu'}S^{\mu} 
 \rangle-\frac{1}{3}S(S+1)\delta_{\mu\mu'}$.\cite{TsuneAri} 
 They investigated an S=1 spin model with bilinear and biquadratic (BLBQ) 
couplings on the triangular lattice, defined as
\begin{equation}
H=\sum_{\langle i,j\rangle}\left[J\bm{S}_i\cdot\bm{S}_j+K(\bm{S}_i
\cdot\bm{S}_j)^2\right]. \label{i1}
\end{equation}
First they showed this model has an AFQ order that fits the tripartite 
triangular lattice in the parameter region of $0<J<K$ 
using mean field approximation. This mean-field ground state is represented as 
\begin{equation}
\ket{\Psi_{MF}}=\prod_{\bm{R}}\ket{S_x=0}_{A,\bm{R}}\otimes
\ket{S_y=0}_{B,\bm{R}}\otimes\ket{S_z=0}_{C,\bm{R}}, \label{mfgs}
\end{equation}
where $j$ labels three sublattices (A, B, and C) and $\ket{S_{\alpha}=0}_{j,\bm{R}}$ 
denotes the single-spin state with $0$ eigenvalue of $S_{\alpha}$-operator ($\alpha=x,y,\ or\ z)$ 
at the $j$-sublattice site in the unit cell $\bm{R}$.
Then they studied low energy 
properties in the AFQ phase using a bosonic description of the excitation and 
obtained results qualitatively consistent with the three essential 
points in the experiments in ${\rm NiGa_2S_4}$: (1) absence of magnetic 
long-range order; (2) nonvanishing susceptibility at zero temperature;
 (3) $T^2$ behavior of the specific heat. 
L\"{a}uchli {\it et al.} independently studied the ferro quadrupolar (FQ) phase 
in the parameter region $K<J\lesssim -2.5K$ of the same BLBQ model and obtained the results 
similar to the AFQ case.\cite{Penc} 
Although these proposals are suggestive, we have to note that the origin of 
 large effective biquadratic coupling, either positive or negative, 
 remains to be clarified. Further, a more direct identification of quadrupolar 
 order is desired.

Spin freezing is another unusual phenomenon observed in ${\rm NiGa_2S_4}$. The 
magnetic susceptibility shows a kink at $T_f=8.5\rm{K}$, and a small 
bifurcation between field cooling (FC) and zero field cooling (ZFC) values 
below $T_f$\cite{Science}. Muon spin rotation ($\mu$ SR) experiments 
revealed quasistatic relaxation of Ni spin below $T_f$.\cite{muSR} 
These results suggest a spin freezing transition at $T_f$. 
However, the characteristic of the spin frozen state below $T_f$ is remarkably 
different from the case of canonical spin glass materials. Slow Ni-spin fluctuations 
with a time scale of ${\rm \mu s}$ exist and are rapidly suppressed upon 
application of magnetic 
field $\geq 10 {\rm mT}$.\cite{muSR} In order to further investigate 
this spin-freezing transition, Nambu {\it et al.} studied 
the thermodynamic properties of ${\rm Ni_{1-{\it x}}Zn_{\it x}Ga_2S_4}$, 
where Ni ions are partially replaced with nonmagnetic Zn ions.\cite{SDIE} 
They showed that the freezing temperature $T_f$ decreases with 
increasing impurity concentration $x$. This is just opposite to the  case of canonical 
spin glass materials. It is also important that $T_f$ scales with Weiss temperature, which is 
also the characteristic energy scale of the low temperature specific 
heat. 

The main purpose of this paper is to propose a novel mechanism of spin freezing 
that is consistent with the spin liquid behavior and the unconventional 
spin freezing in ${\rm NiGa_2S_4}$. Assuming the existence of the AFQ order, we will 
introduce impurity magnetic moments in the system and study interaction between 
them mediated by low energy excitation in the AFQ order. We will then discuss 
a possibility of spin-freezing caused by this interaction.

This paper is organized as follows. In \S 2, we will introduce a model for a single
disorder, which induces magnetic moments in otherwise AFQ ordered system. 
We also describe the strategy of our calculations. 
In \S 3 we will derive effective continuum models to describe 
low-energy excitation in the AFQ order. Using this, we will study the one-impurity 
problem in \S 4, to investigate the coupling between an individual magnetic impurity 
and low-energy excitations in the bulk. In \S 5, we will 
derive interactions between the impurity magnetic moments, mediated by the 
low-energy bulk excitation. Then we will discuss the possibility of spin-freezing 
caused by this interaction in \S 6. Finally \S 7 is a short summary.

\section{Model and Strategy}
\begin{figure}[b]
\centering
\includegraphics[scale=0.44,viewport=0 0 537 155]{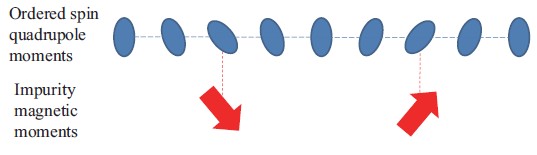}
\caption{Schematic diagram for the interaction between impurity magnetic 
moments. Dotted lines denote abstract interactions.}
\label{rkky}
\end{figure}

\begin{figure}[t]
\centering
\includegraphics[scale=0.44,viewport=0 0 558 325]{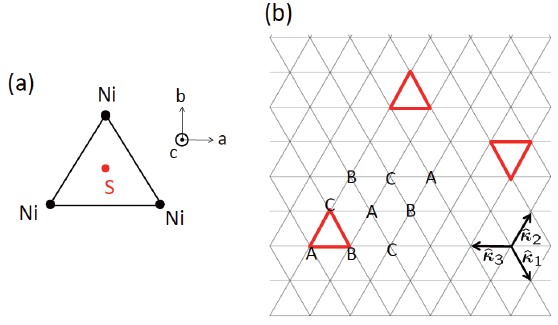}
\caption{(a)Position of a sulfur site in ${\rm NiGa_2S_4}$. (b)Triangle bond 
disorder, three sublattices, and lattice vectors
$\hat{\bm{\kappa}}_i\ (i=1,2,3)$ defined below \eref{eq126}.}
\label{TBD}
\end{figure}

We start with introducing a microscopic model of ${\rm NiGa_2S_4}$ including 
magnetic disorder, which can explain observed spin freezing, in \S2.1. 
We mainly study the case of $T=0$ and our basic assumption is that the system 
has the AFQ order. Then we describe our basic strategy of calculations in \S 2.2.

\subsection{Model}
One feature of the spin freezing in ${\rm NiGa_2S_4}$ is that the freezing 
temperature $T_f$ scales with the characteristic energy scale of low-temperature 
specific heat, which is ascribed to collective excitations in the AFQ 
order. This implies that the spin freezing has close relationship with 
the AFQ order. 
We consider that the freezing is induced by the interaction between 
disorder-induced magnetic moments, which is mediated by low energy excitations in the 
AFQ order, as schematically shown in \figref{rkky}.

As a specific realization of this scenario, we construct a microscopic model 
as follows. First, for the pure bulk system to which impurities are to be introduced, 
we employ the BLBQ model \erefn{i1}, which is a minimal model for 
the AFQ order. 
We investigate the parameter region 
$0<J<K$, where the ground state has the AFQ order. 
Note that the bilinear interaction is rather simplified in this 
model. Large third-neighbor interaction is believed to exist in 
${\rm NiGa_2S_4}$\cite{Science}, but it is not included in our model. 
We use the model \erefn{i1} for simplicity of the discussion.
  Second, we introduce disorders to the system. Here we focus on bond disorder, 
not site disorder, considering experimental results. Small deficiency of sulfur 
concentration drastically enhances the glassy behavior in hysteresis of 
the magnetic susceptibility.\cite{sulfur} Sulfur ions are located above or below the 
center of each triangular plaquette, as
 shown in \figref{TBD}(a), and their orbitals are in dominant exchange pathways.
Disorder is induced in exchange interaction due to the vacancy of sulfur sites, 
and this is 
essential for the spin freezing behavior. Taking this into account, 
we introduce a different bilinear coupling for
 the bonds in a triangular plaquette at the impurity position. Thus, the model 
 reads
\begin{align}
H=&\sum_{\langle i,j \rangle \notin D}^{n.n.}\left[J\bm{S}_i\cdot
             \bm{S}_j+K(\bm{S}_i\cdot\bm{S}_j)^2\right] \notag\\
     &+\sum_{\langle i,j\rangle \in D}^{n.n.}\left[J'\bm{S}_i\cdot
             \bm{S}_j+K(\bm{S}_i\cdot\bm{S}_j)^2\right], \label{g1}
\end{align}
where $D$ denotes randomly distributed triangular plaquettes and 
an example configuration is shown in 
\figref{TBD}(b). Hereafter we call this individual triad of disorder bonds, 
simply {\it impurity}. We assume that the biquadratic coupling 
$K$ in \eref{g1} is not affected by impurities, since we focus on the behavior of the 
magnetic dipole moments induced by disordered exchange couplings, while the 
local variation in $K$ does not yield significant results. Further, we study 
the ferromagnetic case $J'<0$, in which the model is consistent with the scenario above, 
since three spins on an impurity plaquette tend 
to align and form an impurity magnetic moment as a whole. 

\subsection{Basic Strategy}
Before starting calculations, it is useful to describe the framework and limitations of the present study. 
Our goal is to obtain the interaction between impurities in the AFQ ordered state. 
These interactions arise from the interference of the modulations of the AFQ order and they have two parts. 
The first contribution is related to the fact that each impurity deforms the nematic order pattern in 
the host locally around it, and it is given by the interference of this static order parameter deformation between 
the impurities. 
The second contribution is mediated by the interactions of impurities and quantum excitations in the bulk. 
One impurity interacts with different sets of excitations depending on impurity magnetic state. Those 
excitations propagate in the bulk and interact with another impurity, which is also dependent on the 
magnetic state of the second impurity, and this leads to impurity-impurity interactions. 

In the present study, we focus on the first contribution, {\it i.e.} the one given by static deformation of the nematic order 
due to impurity and neglect the contribution of dynamical quantum excitations. This may be partially justified by the 
fact that the AFQ order is stable in the parameter region $(0<J<K)$ of the BLBQ model and the reduction of the nematic order 
parameter due to quantum fluctuations is quite small\cite{TsuneAri, Penc}. This does not exclude the possibility that 
quantum fluctuations play some essential role, but this problem is beyond the scope of this study and should be examined in 
the future. However, based on a heuristic argument, we expect that dynamical quantum effects also lead to impurity-impurity 
interactions with similar nature, and we will discuss this briefly at the end of \S 6. 


To describe static deformations of the nematic order, we employ a site-dependent mean field approximation. 
The phase space is restricted to the subspace of site-factorized wave functions 
 $\ket{\Psi_{\rm MF}}=\prod_i\psi_i$ where $\psi_i$ denotes a one-spin wave function at 
site $i$. 
Local nematic and magnetic order parameters are given by 
$Q^{\mu\nu}_i=\braket{\psi_i|\frac{1}{2}(S_i^{\mu}S_i^{\nu}+S_i^{\nu}S_i^{\mu})|\psi_i}-\frac{1}{3}S(S+1)\delta_{\mu\nu}$ and 
$\bm{m}_i=\braket{\psi_i|\bm{S}_i|\psi_i}$ respectively, and the energy of the corresponding 
configuration is given by $\Braket{\Psi_{\rm MF} | H|\Psi_{\rm MF}}$. 
In this way, the energy is a functional of these local fields $\{Q\}$ and $\{\bm{m}\}$, and will construct 
a "classical" Hamiltonian describing this energy cost and its continuum limit. This is a 
classical model because only static deformations are considered there. 
Low energy configurations within this approximation are accompanied with long-wavelength distortions of 
the order parameters, and this distortion is referred to as excitation in the following. 

Within this framework, we will evaluate the energy and configuration 
of the ground state with two impurities. The result shows peculiar nature of impurity-impurity interaction, 
and this interaction is a key of novel type of spin freezing, 
which can describe the peculiarity of the spin freezing phenomena observed in ${\rm NiGa_2S_4}$. 

\section{Continuum Theory of the Bulk}
The bulk part of the model \erefn{g1} behaves as a medium of the interaction between impurities 
and only low energy excitations in the AFQ order play a significant role, 
while detailed lattice structure is not important. 
It justifies replacing the bulk part with an effective field theory describing 
low energy excitation, and let us derive the effective model in this section. 
There are two kinds of excitations. One is nonmagnetic excitation 
corresponding to deformation of the order of spin quadrupole moments. The other 
is magnetic excitation, which induces magnetic dipole moments. We introduce field 
variables describing these excitations and derive effective models up to the 
second order in these fields. Up to this order, nonmagnetic and magnetic excitations 
are decoupled. 

It is convenient to introduce the following time-reversal invariant basis for each site:
\cite{Ivanov}
\begin{equation}
\ket{x}=i\frac{\ket{1}-\ket{\overline{1}}}{\sqrt{2}},\ \ket{y}=\frac{\ket{1}+
       \ket{\overline{1}}}{\sqrt{2}},\ \ket{z}=-i\ket{0}, \label{eq3}
\end{equation}
where $\ket{1}, \ket{0},$ and $\ket{\overline{1}}$ denote the eigenstates of 
 $S_z$ operator with eigenvalues $1,0,$ and $-1$ respectively. 
 A general one-spin wave function is represented as
\begin{equation}
\ket{\bm{d}}=d_x\ket{x}+d_y\ket{y}+d_z\ket{z},\quad \bm{d}\in\bm{\mathbb{C}}^3 
  \label{eq4}
\end{equation}
and we define two real vectors as the real and imaginary parts of $\bm{d}$:
\begin{equation}
\bm{d}=\bm{u}+i\bm{v},\qquad \bm{u},\bm{v}\in\bm{\mathbb{R}}^3.
\label{eq5}
\end{equation}
These vectors satify the normalization condition$|\bm{u}|^2+|\bm{v}|^2=1$, 
and can also satisfy the orthogonality $\bm{u}\cdot\bm{v}=0$ and the condition 
$|\bm{u}|\geq|\bm{v}|$ by choosing an appropriate phase factor. We choose a 
local phase satisfying these relations 
 in the rest of this paper. Using this representation, the expectation value of 
 the Hamiltonian 
 \erefn{i1} with regard to the site-decoupled wave function 
 $\ket{\Psi_{\rm MF}}=\prod_i\ket{\bm{d}_i}$ is written as
\begin{align}
&\Braket{\Psi_{\rm MF} | H|\Psi_{\rm MF}}= \nonumber \\
&\sum_{\langle ij\rangle}\left\{(4J-2K)[(\bm{u}_i\cdot\bm{u}_j)(\bm{v}_i
\cdot\bm{v}_j)-(\bm{u}_i\cdot\bm{v}_j)(\bm{v}_i\cdot\bm{u}_j)] \right.\nonumber\\
&\left. +K[(\bm{u}_i\cdot\bm{u}_j)^2+(\bm{v}_i\cdot\bm{v}_j)^2+
 (\bm{u}_i\cdot\bm{v}_j)^2+(\bm{v}_i\cdot\bm{u}_j)^2]\right\}.
\label{eq9}
\end{align}
On the basis of this expression, we will derive the effective model for nonmagnetic excitation 
in \S 3.1 and then will turn to the effective model for magnetic excitation in \S 3.2.

\subsection{Effective Model for Nonmagnetic Excitation}
Let us derive the effective model for nonmagnetic excitation first. 
This describes the energy of configurations under the condition that the magnetic moment $\bm{m}=0$ 
at any site. We will show the effective model is the ${\rm O(4)}$ nonlinear-$\sigma$ model.
Since the magnetic moment is given by
 $\bm{m}=\langle\bm{S}\rangle=2\bm{u}\times\bm{v}$, the condition $\bm{m}=0$ 
 corresponds to $|\bm{u}|=1$ and $\bm{v}=0$. In this case one-spin state is 
 characterized by vector $\bm{u}$, referred to as {\it director}. Note that 
 this representation is double-valued. Two states with directors $\pm\bm{u}$ 
 differ only by a phase factor and hence correspond to the identical physical
state.
  The Hamiltonian \erefn{eq9} becomes
\begin{equation}
H=\sum_{\langle ij\rangle}K(\bm{u}_i\cdot\bm{u}_j)^2. \label{eq23}
\end{equation}

In the ground state of the disorder-free bulk system, directors in each sublattice are 
spatially uniform and orthogonal between different sublattices. 
Nonmagnetic excitations mean a long-wavelength distortion of this set of ordered 
directors. Hence, as a local order parameter, we can use an orthogonal triad 
of unit vectors, which corresponds 
to the directors at mutually nearest neighbor sites belonging to the three 
sublattices. Strictly speaking, the orthogonality of the directors between 
nearest neighbor sites can be slightly violated, but we can ascribe this 
deviation to spatial variation of the local order parameter. 
We represent this triad as a ${\rm SO(3)}$, or almost equivalently a ${\rm SU(2)}$ rotation operation, 
and here we choose the latter for convenience.
This representation has multivalueness since the directors are double-valued 
and also each ${\rm SO(3)}$ matrix has double representation in ${\rm SU(2)}$. 
Although such redundancy plays an essential role for topological excitation, 
it is not relevant to the effects of spin-wave like excitation, which is 
studied in this paper. (We will summarize the properties of topological excitation 
in this system in Appendix B, and the important point is that the non-Abelian 
fundamental group of the AFQ order parameter implies a nontrivial merging rule 
of topological excitations. ) 
Therefore, we hereafter use the spin-$1/2$ representation of the ${\rm SU(2)}$ group, i.e. 
${\rm SU(2)}$ matrix.
Due to the locality and the ${\rm SU(2)}$ invariance, 
we can expect the effective 
model will have a following form, 
\begin{equation}
H^{\pi}=\frac{J_{\pi}}{2}\int d\bm{r}{\rm Tr}\left[\nabla U^{\dag}(\bm{r})\cdot\nabla U(\bm{r})\right], \label{nls}
\end{equation}
with ${\rm SU(2)}$ field variable $U(\bm{r})$ and stiffness constant $J_{\pi}$. 
After a standard parametrization, this model is mapped to the ${\rm O(4)}$ nonlinear-$\sigma$ model. 
Let us now derive this effective continuum model from 
the lattice model \erefn{eq23} and also determine the value of $J_{\pi}$. 
We construct the correspondence between the triad and the matrix as follows. 
First we define a double-valued unit 
 vector $\bm{\xi}(\bm{x}_i),\ \bm{\eta}(\bm{x}_i),{\textrm or}\ 
   \bm{\zeta}(\bm{x}_i)$ depending on sublattice as
\begin{align}
&\bm{\xi}(\bm{x}_i)=\pm \bm{u}_i\quad i\in A,\ 
\bm{\eta}(\bm{x}_i)=\pm \bm{u}_i\quad i\in B,\nonumber \\ 
&\bm{\zeta}(\bm{x}_i)=\pm \bm{u}_i\quad i\in C.
\label{eq24}
\end{align}
Under appropriate choices of signs, all vectors vary in space slowly 
compared with the lattice constant in the low energy sector.
Then we extend these vectors by interpolation to those defined for continuous spatial 
coordinate $\bm{r}$ and impose the orthogonality
 among the three fields.
Lastly we define the ${\rm SU(2)}$ field variable $U(\bm{r})$ by
\begin{align}
&U(\bm{r})\bm{\chi}[{\bm{x}}]=\bm{\chi}[{\bm{\xi}(\bm{r})}],\  
U(\bm{r})\bm{\chi}[{\bm{y}}]=\bm{\chi}[{\bm{\eta}(\bm{r})}],\nonumber\\
&U(\bm{r})\bm{\chi}[{\bm{z}}]=\bm{\chi}[{\bm{\zeta}(\bm{r})}],  \label{eq26}
\end{align}
where $\bm{x}=(1,0,0),\ \bm{y}=(0,1,0),\ \bm{z}=(0,0,1)$, and $\bm{\chi}[\bm{N}]$ 
denotes the two component spinor representation of the spin 1/2 coherent state 
\cite{Sachdev} corresponding to the unit vector $\bm{N}$:
\begin{equation}
\bm{\chi}[\bm{N}]=\left(
\begin{array}{c}
\cos\left(\frac{\theta}{2}\right)\exp\left(-\frac{i\phi}{2}\right) \\
\sin\left(\frac{\theta}{2}\right)\exp\left(+\frac{i\phi}{2}\right),
\end{array}
\right) \label{eq26.2}
\end{equation}
using polar coordinates as 
$\bm{N}=(\sin\theta\cos\phi,\sin\theta\sin\phi\\,\cos\theta)$. 
It is known that a ${\rm SU(2)}$ matrix is written with four real parameters as 
\begin{align}
&U(\bm{r})=\pi_0(\bm{r})+\sum_{j=1}^3i\tau_j\pi_j(\bm{r})\nonumber\\
&\pi_{\mu}\in\mathbb{R},\quad(\mu=0,1,2,3),\quad \sum_{\mu=0}^3 \pi_{\mu}^2=1,
\label{eq32}
\end{align}
where $\tau_j$'s$(j=1,2,3)$ are Pauli matrices. Hereafter we regard $\pi_{\mu}$
's $(\mu=0,1,2,3)$ as local order parameters.

Next we rewrite the bond Hamiltonian \erefn{eq23} in terms of the field 
variables $\pi_{\mu}(\bm{r})$. We can derive straightforwardly the 
representation of the continuated director vectors, $\bm{\xi}(\bm{r})$, 
$\bm{\eta}(\bm{r})$, and $\bm{\zeta}(\bm{r})$, such as
\begin{align}
\bm{\xi}=&\left(\pi_0^2+\pi_1^2-\pi_2^2-\pi_3^2,2(\pi_1\pi_2-\pi_0\pi_3),\right.
                                                  \nonumber\\
& \left.2(\pi_0\pi_2+\pi_1\pi_3)\right). \label{eq43}
\end{align}
Using these relations, and rewriting $\bm{\pi}(\bm{x}_i)$ and 
$\bm{\pi}(\bm{x}_j)$ as $\bm{\pi},$ and $\bm{\pi}+\delta\bm{\pi}$ respectively,
we have interaction
\begin{align}
H_{ij}&=4K\left(-\pi_3\delta\pi_0-\pi_2\delta\pi_1+\pi_1\delta\pi_2
              +\pi_0\delta\pi_3\right)^2 \nonumber\\
&\equiv H_{AB}(\bm{r},\bm{r}+\delta\bm{r}),
\label{eq49.5}
\end{align}
between the sites $i$ and $j$ when $i\in A,\ j\in B$. We similarly define 
$H_{BC}$ and $H_{CA}$.

Summing over all the site pairs and taking the continuum limit, we obtain the 
${\rm O(4)}$ nonlinear-$\sigma$ model as the effective model for nonmagnetic excitations, 
as we expected in \eref{nls}.
\begin{equation}
H^{\pi}=\frac{1}{2}J_{\pi}\sum_{\mu=0}^3\int d\bm{r} (\nabla\pi_{\mu})^2,\qquad 
\sum_{\mu=0}^3\pi_{\mu}^2=1, \label{eq62}
\end{equation}
where $J_{\pi}=(8/\sqrt{3})K$. As shown by Polyakov and Wiegmann, this model is identical to 
the phenomenologically introduced effective model \erefn{nls}\cite{PW}, and the value of the coupling 
constant $J_{\pi}$ is determined explicitly. 
This classical model describes 
long-wavelength distortion of the AFQ order, 
which physically corresponds to 
the energy cost of instantaneous deformation\cite{model_comment}. 
Note that the order parameter space ${\rm SU(2)}$ reflects the character of 
antiferro order. Due to the orthogonality of directors between 
different sublattices, there are locally three independent 
directions of distortion of the order, which correspond to rotations of 
ordered moments around three axes in the spin space. 
On the other hand, in the ferro quadratic order where directors point to the 
same direction regardless of sublattice, with the rotation around the director
the order remains unchanged and the number of independent directions of distortion 
is two. Therefore we can easily conclude the effective model for the 
ferro quadratic order is ${\rm O(3)}$ nonlinear-$\sigma$ model.

Finally we introduce further simplification for the effective model. We 
describe the ground state of the pure bulk as $\pi_0=1,\ \pi_a=0,\ (a=1,2,3)$. 
Even in the presence of an impurity, the bulk region is 
close to this ground state. Hence we assume
 $|\pi_a|\ll \pi_0\sim 1,\ (a=1,2,3).$ Expanding the effective model \erefn{eq62} 
 in terms of the field variables $\pi_a\ (a=1,2,3)$ and 
preserving only the lowest order terms of $\pi_a$, we obtain  the three-component massless
Gaussian model:
\begin{equation}
H^{\pi}=\frac{1}{2}J_{\pi}\sum_{a=1}^3\int d\bm{r}(\nabla\pi_{a})^2 \label{z3}
\end{equation}

Note that contribution of $\pi_0$ is $O(\pi_a^4)$ and therefore neglected here. 
We will use this effective model to study nonmagnetic excitations afterwards. 
This simplification of the model is not appropriate for investigating 
the excitations where $\pi-$field configuration strongly deviates from that 
in the ground state. For example, at finite temperatures, the long range order 
is destroyed as manifested by Mermin-Wagner theorem\cite{MW}, and the 
$\pi-$field varies in space among the whole parameter space even without 
disorders. To deal with such a situation, we have to go back to the original 
model \erefn{eq62}.

\subsection{Effective Model for Magnetic Excitation}
Now we turn to derive an effective model for magnetic excitation, which 
will be identified as a three-component massive Gaussian model. 
In contrast to its nonmagnetic partner, this describes the 
energy of configuration without nonmagnetic excitation.
 This condition means that the directions of the directors $\bm{u}$ do not 
 change from those in the bulk ground state, while $|\bm{u}|$ can be 
 smaller than the bulk value, $|\bm{u}|=1$. Hence we can impose
\begin{equation}
\bm{u}_i\sslash \bm{x} \quad (i\in A),\   
\bm{u}_i\sslash \bm{y} \quad (i\in B),\  
\bm{u}_i\sslash \bm{z} \quad (i\in C).  \label{eq63.3}
\end{equation}

First let us introduce field variables describing magnetic excitation. 
Basically we can choose three components of magnetic moments as the field variables 
in concern, 
but we have to pay attention to the following two points. 
First, condition \erefn{eq63.3} yields a restriction of magnetic moments.
For example, in the A sublattice, the magnetic moment is restricted in the YZ plane, 
$\bm{m}=\left(0,m_2^{(A)},m_3^{(A)}\right)$, since $\bm{m}=2\bm{u}\times\bm{v}$. 
Second, coupling between 
magnetic moments is ferromagnetic or antiferromagnetic, depending 
on the parameter $J/K$. 
It is clarified by transforming the total Hamiltonian \erefn{eq9} to the form
\begin{align}
&\langle H_{ij}\rangle=\left(J-\frac{1}{2}K\right)\bm{m}_i\cdot\bm{m}_j 
            \nonumber\\
    &+K\{(\bm{u}_i\cdot\bm{u}_j)^2+(\bm{v}_i\cdot\bm{v}_j)^2+
(\bm{u}_i\cdot\bm{v}_j)^2+(\bm{v}_i\cdot\bm{u}_j)^2\}. \label{eq20.9}
\end{align}
The interaction between $\bm{m}_i$ and $\bm{m}_j$ is ferromagnetic 
(antiferromagnetic) when $0<J<K/2$ ($K/2<J<K$). 
Taking these points into account, we choose field variables as follows. 
In the ferromagnetic region $0<J<K/2$, the three variables are
defined as $\psi_1=\frac{1}{2}m_1^{(B)}=\frac{1}{2}m_1^{(C)}$, 
$\psi_2=\frac{1}{2}m_2^{(C)}=\frac{1}{2}m_2^{(A)}$, and 
$\psi_3=\frac{1}{2}m_3^{(A)}=\frac{1}{2}m_3^{(B)}$ 
(here the subscript denotes a component, not a site), which correspond to "uniform" 
magnetizations, as the field variables varying gradually in space. 
In contrast, in the antiferromagnetic region $K/2<J<K$, the field 
variables are 
$\psi_1=\frac{1}{2}m_1^{(B)}=-\frac{1}{2}m_1^{(C)}$,
$\psi_2=\frac{1}{2}m_2^{(C)}=-\frac{1}{2}m_2^{(A)}$, and 
$\psi_3=\frac{1}{2}m_3^{(A)}=-\frac{1}{2}m_3^{(B)}$
, which represent "staggered" magnetizations. 

Next we express the Hamiltonian in terms of the field variables $\psi$ and derive an 
effective model. Using the field variables defined above, the Hamiltonian 
\erefn{eq9} is written as:
\begin{equation}
H=\sum_{ i\in A,j\in B,k\in C}^{\rm n.n.}\left( H_{ij}+H_{jk}+H_{ki} \right) 
            \label{eq75.9} 
\end{equation}
\begin{equation}
H_{ij}=-|4J-2K|\psi_{3,j}\psi_{3,i}+K(\psi_{3,j}^2+\psi_{3,i}^2)\  
      \textrm{etc.} \label{eq72} \\
\end{equation}

As in the previous discussion for the nonmagnetic part,
we take the continuum limit, and the result is the three-component
Gaussian model with mass terms:
\begin{equation}
H^{\psi}=\frac{1}{2}J_{\psi}\sum_{a=1}^3\int d\bm{r}
        \left[(\nabla\psi_a)^2+(k_{\psi})^2\psi_a^2\right], \label{eq83}
\end{equation}
where the coupling constants are given by
\begin{equation}
J_{\psi}=\frac{2}{\sqrt{3}}|K-2J|,\ k_{\psi}=\frac{2}{l_0}\sqrt{\frac{K}{|K-2J|}-1}, 
                   \label{eq84}
\end{equation}
where $l_0$ denotes the lattice constant. Hereafter we take $l_0=1$ for simplicity. 
The effective model \erefn{eq83}, is {\it massive}, reflecting short-range nature 
 of magnetic correlations of the AFQ order. 
Note that the "mass" $k_{\psi}$ vanishes as we approach the boundaries of the 
AFQ phase, $J=0,$ or $K$. This singularity implies that the system becomes 
unstable against magnetic excitation, and manifests the 
onset of a magnetically ordered state. 
Actually, the mean-field ground state has ferromagnetic order for $J<0$, 
and 120 degree antiferromagnetic order for $J>K$\cite{Penc}.

\section{One Impurity Problem}
In this paper, we want to calculate interaction between two impurities in the AFQ order, 
and this interaction is mediated by coupling of each impurity and bulk excitations. 
Therefore, our next task is to study the problem of single impurity and calculate its 
coupling constant of the interaction with nonmagnetic and magnetic excitations in the 
bulk studied in the last section. 
The most important 
result is that induced nonmagnetic excitation field shows a power-law decay in space, much 
more extended than magnetic excitation field.

Our strategy is as follows. For the bulk part we use the continuum theory 
developed in the last section, while we use the original spin variables for 
the impurity part. We then derive the bulk-impurity coupling by evaluating 
the energy on the bonds connecting these two parts.

\begin{figure}[!t]
\centering
\includegraphics[scale=0.44,viewport=0 0 568 278]{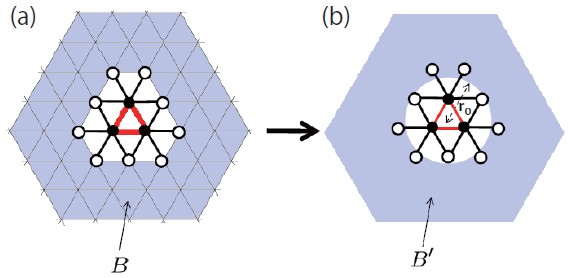} 
\caption{Simplification of the one-impurity 
Hamiltonian. (a) Black (white) circles denote core (shell) sites. (b)The bulk 
is replaced by a continuum media and its inner boundary is approximated by a circle.}
\label{bbh_approx}
\end{figure}

At this point we explain our nomenclature concerning the sites and bonds. 
We denote sites in the impurity as "core" sites, and sites on the border of the bulk 
which are connected with core sites by bonds as "shell" sites. 
They are shown as black and white circles respectively in 
\figref{bbh_approx}(a). We call the other sites "bulk" sites. Further we call 
bonds connecting core sites and shell sites (two core sites) "core-shell" 
("core-core") bonds, which correspond to red (black) bonds in 
\figref{bbh_approx}(a).
Finally we call the other bonds "bulk" bonds, which are shown as gray bonds 
in the figure. 

First, we represent states on shell and bulk sites 
using the two vector field variables $\bm{\pi}$ and $\bm{\psi}$ defined in the last 
section, instead of original spin wavefunctions.
For core sites, we keep using original spin wavefunctions.
Next, we evaluate the energy of bulk bonds using the effective massless and 
massive Gaussian models derived in the last section, whereas 
energy of core-core and core-shell bonds are evaluated for the original 
lattice model \erefn{i1}. In doing this, spin wavefunction at a shell 
site is determined by the field variables at its position.
Then, we approximate the boundary of the bulk part, which has a 
polygon shape in the original lattice model, by a circle of radius $r_0$,
 to make it easier to evaluate the energy of the bulk part. 
Finally, we consider only the dominant components of spatial fluctuations of 
the fields.

For simplicity, we take the limit $J'\rightarrow -\infty$ i.e. the magnetic 
interaction between core sites is ferromagnetic one of infinite strength. 
In this limit, spins on three core sites are completely polarized and described 
by the identical magnetic moment vector of unit length 
$\bm{m},\ |\bm{m}|=1$. We change $\bm{m}$ and clarify the anisotropy of 
the magnetic moment. We will show later, using numerical calculations, that 
the cases of finite  $J'\ (J'<0)$ are qualitatively similar to this limit.

The total Hamiltonian is 
\begin{equation}
H=H_{\rm bulk}^{\pi}+H_{\rm bulk}^{\psi}+H_{\rm imp}, \label{a1}
\end{equation}
where $H_{\rm bulk}^{\pi}$ and $H_{\rm bulk}^{\psi}$ are the effective 
Hamiltonians for the nonmagnetic and magnetic excitations \erefn{z3} 
and \erefn{eq83}, and $H_{\rm imp}$ denotes the bond Hamiltonians on core-shell 
bonds. $H_{\rm imp}$ represents impurity-bulk interaction and 
is a function of the magnetic moment $\bm{m}$ of core 
sites and the field variables $\bm{\pi}$ for nonmagnetic excitation 
and $\bm{\psi}$ for magnetic excitation on shell sites.

First we will derive $H_{\rm imp}$ perturbatively with regard to $\bm{\pi}$ and $\bm{\psi}$ 
up to the second order in \S 4.1. Then, we will consider only the first order term within 
$H_{\rm imp}$ and examine the magnetic anisotropy. 
At this order we can decouple the total Hamiltonian \erefn{a1}
into nonmagnetic part $H^{\pi}$ and magnetic part $H^{\psi}$.  
We will treat them separately in \S 4.2 and \S 4.3, respectively. 
Next we will further include the second 
order term and refine the anisotropy. Since we can not decouple the Hamiltonian as before, we will treat 
the whole Hamiltonian of this order in \S 4.3.
Finally we will approach this problem numerically 
and compare the results with analytical ones in \S 4.4.

\subsection{Perturbative Expansion of the Impurity-Bulk Interaction}
Following the strategies mentioned above, firstly we expand the impurity-bulk 
interaction $H_{\rm imp}$ with regard to the two field variables, $\bm{\pi}$ and $\bm{\psi}$. 
An individual 
bond Hamiltonian $H_{{\rm imp},j}$, connecting the core site $i$ and the shell site 
$j$, is written in terms of the impurity magnetic moment $\bm{m}$, and 
the field variables on $j$ site: $\bm{\pi}_j$ and $\bm{\psi}_j$, up to the second order 
of $\bm{\pi}_j$ and $\bm{\psi}_j$. 
Here we explain for the case where the site $j$ belongs to A sublattice. 

If there is no nonmagnetic excitation at a shell site, we can derive the 
correspondence between the two vectors ($\bm{u}$ and $\bm{v}$) and the field of 
magnetic excitation $\bm{\psi}$, from the definition of $\bm{\psi}$.
As mentioned before, nonmagnetic excitation corresponds to rotation of local 
nematic order in the spin space. The directions of principal axes 
are given by the representation of the directors, \eref{eq43}, from 
which we can identify the rotation matrix. Thus we represent 
$\bm{u}$ and $\bm{v}$, in terms of both magnetic and nonmagnetic excitation:
\begin{align}
\bm{u}=&\left(1-2(\pi_1^2+\pi_2^2)-\right.\frac{1}{2}(\psi_3^2-\sigma\psi_2^2),
                                  \nonumber\\
       & \left.\quad 2(\pi_1\pi_2-\pi_3),2(\pi_2+\pi_1\pi_3)\right) \label{eq88} \\
\bm{v}=&\left(2\pi_3\psi_3-2\pi_2\sigma\psi_2,\psi_3-2\pi_1\sigma\psi_2,-
                                \sigma\psi_2-2\pi_1\psi_3\right), \label{eq89}
\end{align}
where the third and higher order terms are omitted and 
$\sigma=+1\ (-1)$ for $J<K/2\ (J>K/2)$. 
As for the core site $i$ with magnetic moment $\bm{m}$, we can readily obtain 
the value of $\bm{u}$ and $\bm{v}$, from the relations among these three 
vectors: 
$|\bm{u}|^2+|\bm{v}|^2=1,\ \bm{u}\cdot\bm{v}=0,$ and $\bm{m}=2\bm{u}\times\bm{v}$. 
Combining these results with the bond  Hamiltonian in \eref{eq9}, we obtain the 
representation of $H_{{\rm imp},j}(\bm{m},\bm{\pi}(\bm{x}_j),
\bm{\psi}(\bm{x}_j))$ for $j\in A$.
\begin{align}
H_{{\rm imp},j}=&H_{{\rm imp},j}^{\pi}+H_{{\rm imp},j}^{\psi}+
   H_{{\rm imp},j}^{\pi^2}+H_{{\rm imp},j}^{\psi^2}+H_{{\rm imp},j}^{\pi\psi}
     \label{eq91} \\
H_{{\rm imp},j}^{\pi}=&2K\epsilon_{1ab}m_1m_a\pi_b(\bm{x}_j) \label{eq92} \\
H_{{\rm imp},j}^{\psi}=&(2J-K)\left[-\sigma m_2\psi_2(\bm{x}_j)+m_3
                                     \psi_3(\bm{x}_j)\right] \label{eq97} \\
H_{{\rm imp},j}^{\pi^2}=&2K\left\{m_1^2\left[\bm{\pi}^2(\bm{x}_j)-
          \pi_1^2(\bm{x}_j)\right]-\left[\epsilon_{1ab}m_a\pi_b
                                (\bm{x}_j)\right]^2 \right\} \label{eq94} \\
H_{{\rm imp},j}^{\psi^2}=&\frac{K}{2}\left\{m_1^2\left[\bm{\psi}^2
           (\bm{x}_j)-\psi_1^2(\bm{x}_j)\right]-\left[\epsilon_{1ab}
            m_a\psi_b(\bm{x}_j)\right]^2\right\} \label{eq98} \\
H_{{\rm imp},j}^{\pi\psi}=&2(2J-K)\left[m_1\epsilon_{1ab}^{(\sigma+1)/2}
                   \pi_a(\bm{x}_j)\psi_b(\bm{x}_j)+ \right. \nonumber \\
             &\left.\pi_1\epsilon_{1ab}m_a\psi_b(\bm{x}_j) \right], \label{eq99}
\end{align}
where $\epsilon_{ijk}$ is Levi-Civita tensor, and summation with regard to 
repeatedly appearing indices is implicitly taken. 
This derivation for the other sublattices
 is straightforward and we obtain the results by changing the indices of 
 $m,\ \pi,$ and $\psi$ in a cyclic way: 
 $1\rightarrow 2,\ 2\rightarrow 3,\ 3\rightarrow 1,\ (j\in B),\ 
1\rightarrow 3,\ 2\rightarrow 1,\ 3\rightarrow 2,\ (j\in C).$ Finally 
the total contribution of all the shell sites is
\begin{equation}
H_{\rm imp}^I=\sum_{j\in {\rm shell}} H_{{\rm imp},j}^I 
     \qquad(I=\pi,\psi,\pi^2,\psi^2,\pi\psi). \label{eq194.4}
\end{equation}

\subsection{First Order Effect of the Interaction between the Impurity and 
   Nonmagnetic Excitation}
At this point we focus on the coupling between the impurity magnetic moment and 
nonmagnetic excitation in the bulk region.
In the impurity-bulk interaction $H_{\rm imp}$, we examine here only the first order term with 
regard to nonmagnetic excitation $\bm{\pi}$, given by \eref{eq92}. 
The total Hamiltonian becomes
\begin{align}
H^{\pi}=&H_{\rm bulk}^{\pi}+H_{\rm imp}^{\pi},\ 
H_{\rm bulk}^{\pi}=\frac{1}{2}J_{\pi}\sum_{a=1}^3\int'd\bm{r}(\nabla\pi_a)^2,\label{eq108} 
\end{align}
where $\int'$ denotes integration over the bulk region $B'$ with a circular void 
shown in \figref{bbh_approx}, 
i.e. $\int' d\bm{r}=\int_{r_0}^{\infty}dr\int_0^{2\pi} d{\theta}$.

As mentioned before, we consider only the dominant component of the spatial fluctuation 
of $\bm{\pi}$. This means as follows: As is clear from $H_{\rm bulk}^{\pi}$ \erefn{z3}, 
the ground state configuration of $\bm{\pi}$ satisfies 
the Laplace equation $\nabla^2\pi_a=0,\ (a=1,2,3)$, and therefore can be 
 expanded in the polar coordinate system as
\begin{equation}
\pi_a(r,\theta)=\sum_{n=1}^{\infty}\left\{c_{n}^{(a)}r^{-n}
           \cos [n(\theta-\theta_{0,n}^{(a)})]\right\}, \label{eq111}
\end{equation}
where we impose the boundary condition $\pi_a(\bm{x})\rightarrow 0$ with 
$|\bm{x}|\rightarrow\infty$. The dipole component $c_1^{(a)}$ becomes dominant
 as $|\bm{x}|\rightarrow\infty$, and we neglect the other components 
 $c_n^{(a)},\ (n>1)$.  Hereafter we call this simply "dipole approximation". 
In this approximation, the field variable $\bm{\pi}$ is given by 
\begin{equation}
\pi_a(r,\theta)\equiv \frac{r_0\bm{\mu}_a^{\pi}\cdot\bm{r}}{r^2}, \label{eq114}
\end{equation}
where we introduced a dipole moment $\bm{\mu}_a^{\pi}$ and its value will be 
determined afterward. Recall that $r_0$ is the radius of the circular void of the 
bulk region. The bulk part of the ground state energy becomes
\begin{equation}
E_{\rm bulk}^{\pi}=\frac{1}{2}\pi J_{\pi}\sum_a(\bm{\mu}_a^{\pi})^2. \label{eq121}
\end{equation}

\begin{figure}[b]
\centering
\includegraphics[scale=0.44,viewport=0 0 293 280]{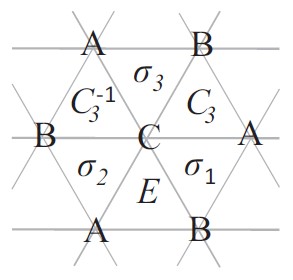}
\caption{Six distinct possibilities of the position of the impurity. 
The position "$E$" is chosen as reference, while the other cases may be 
reproduced by applying a point group operation shown in the figure.}
\label{unit_cell}
\end{figure}

Let us study the ground state of the total Hamiltonian \erefn{eq108}, in the 
dipole approximation. To be specific, we consider the case that the impurity 
is located at position $E$ in \figref{unit_cell}.
The other cases will be summarized in Appendix A.
By straightforward calculation, 
we find that the energy of the lattice part Hamiltonian $H_{\rm imp}^{\pi}$ 
has a simple form:
\begin{equation}
E_{\rm imp}^{\pi}=-\frac{48}{7}K\sum_{a,b,c=1}^3\frac{|\epsilon_{abc}|}{2}
             m_bm_c\hat{\bm{\kappa}}_a\cdot\bm{\mu}_{a}^{\pi}, \label{eq126} 
\end{equation}
where $\hat{\bm{\kappa}}_a$'s are lattice vectors:
$\hat{\bm{\kappa}}_1=(1/2,-\sqrt{3}/2),\ \hat{\bm{\kappa}}_2
                 =(1/2,\sqrt{3}/2),\ \hat{\bm{\kappa}}_3=(-1,0),$ 
shown in \figref{TBD}(b).

The total energy is the sum of the bulk part \erefn{eq121} and the impurity 
part \erefn{eq126}:
\begin{equation}
E=\frac{\pi J_{\pi}}{2}\sum_a(\bm{\mu}_a^{\pi})^2-\frac{48}{7}K\sum_{abc}
   \frac{|\epsilon_{abc}|}{2}m_bm_c\hat{\bm{\kappa}}_a\cdot\bm{\mu}_a^{\pi}
\end{equation}
The dipole moments are determined by minimizing this energy 
with respect to $\bm{\mu}_a^{\pi}$ and the result is
\begin{equation}
\bm{\mu}_{a}^{\pi}=\mu_0^{\pi}\sum_{b,c}\frac{|\epsilon_{abc}|}{2}
                          m_bm_c\hat{\bm{\kappa}}_{a}, \label{eq134}
\end{equation}
where $\mu_0^\pi={48Kr_0}/(7\pi J_{\pi})$. The ground-state energy is 
\begin{equation}
E_0^{\pi}(\bm{m})=\frac{1}{4}\pi J_{\pi}(\mu_0^{\pi})^2
                   \left(m_1^4+m_2^4+m_3^4-1\right). \label{eq137}
\end{equation}

This expression reveals that the impurity magnetic moment has an anisotropy. 
There are four easy axes $\frac{1}{\sqrt{3}}(1,\pm 1,\pm 1)$.
Note that these directions should be considered relative to the principle 
    axes of the local AFQ order, not to the 
spatial directions of the triangular lattice.
Since the Hamiltonian \erefn{g1} is ${\rm SU(2)}$ invariant, we can interpret this 
anisotropy as a result of ${\rm SU(2)}$ symmetry breaking in the AFQ phase. 

\begin{figure}[!t]
\centering
\includegraphics[scale=0.44,viewport=0 0 582 330]{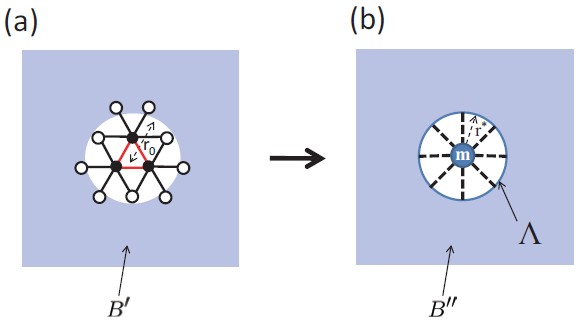}
\caption{Further simplification of the one-impurity 
Hamiltonian. Couplings between core (black) and shell (white) sites are 
replaced by the coupling between the impurity magnetic moment and the boundary 
$\Lambda$ of the bulk region. }
\label{2body_approx}
\end{figure}
 The anisotropy energy \erefn{eq137} can also be represented as
\begin{equation} 
E_0^{\pi}(\bm{M})=-2\pi J_{\pi}(\mu_0^{\pi})^2(M_{xy}^2+M_{yz}^2+M_{zx}^2), \label{q1}
\end {equation}
using $t_{2g}$ part of spin quadrupole moment of each impurity spin
\begin{equation} 
M_{\alpha\beta}\equiv\frac{1}{2}\langle S^{\alpha}S^{\beta}+
  S^{\beta}S^{\alpha}\rangle\quad(\alpha,\beta=x,y,z,\ \alpha\neq\beta),  \label{q2}
\end{equation}
since $M_{\alpha\beta}=m_{\alpha}m_{\beta}/2$ is satisfied for a fully polarized spin 
with $S=1$. 
We can naturally understand this by noticing that the nonmagnetic excitation 
$\bm{\pi}$ linearly couples to the spin quadrupole moment $M_{\alpha\beta}$ in 
the first order coupling \erefn{eq92} considered here. 

We expect the results \erefn{eq137} and \erefn{q1} also explain the magnetic anisotropy 
for finite $J'<0$, although the prefactors may be renormalized. In this case the magnetic moment 
$\bm{m}$ and the spin quadrupole moment $\bm{M}$ should be interpreted as the average among three 
impurity sites. This expectation is consistent with the numerical result, which will be shown in \S 4.5,
that the anisotropy energy for finite $J'<0$ as a function of $J/K$ is qualitatively similar to those for $J'=-\infty$. 

Before closing this subsection, we show for later use that it is possible to modify 
the single-impurity Hamiltonian 
\erefn{eq108} so that the exact ground state coincides with the result of the 
dipole approximation \erefn{eq114} and \erefn{eq134}. 
Detailed structure of lattice part is expected to be irrelevant after the 
dipole approximation and  we modify the Hamiltonian such that the single impurity 
magnetic moment interacts with a field on the boundary $\Lambda$, a circle of 
radius $r^*$, of the bulk region, as shown in \figref{2body_approx}(b). 
Under this simplification, the total Hamiltonian can be written as
\begin{align} 
H^{\pi}=&H_{\rm bulk}^{\pi}+H_{\rm imp}^{\pi} \label{eq248} \\
H_{\rm bulk}^{\pi}=&\frac{1}{2}J_{\pi}\sum_{a=1}^3\int''d\bm{x}(\nabla\pi_a)^2 \label{eq249} \\
H_{\rm imp}^{\pi}=&\int_{\Lambda}d\bm{x}f\left(\bm{m},\bm{\pi}(\bm{x})\right), \label{eq250}
\end {align}
where $\int''$ denotes the integration inside the bulk region $B''$ shown in 
\figref{2body_approx}(b). We require that the ground state of this Hamiltonian 
for impurity magnetic moment $\bm{m}$ is 
given by the result of dipole approximation 
\erefn{eq114} and \erefn{eq134}, which we denote as $\pi_a^0(\bm{m})$.

We can uniquely determine the Hamiltonian which satisfies these conditions:
\begin{equation}  
H^{\pi}=\frac{1}{2}J_{\pi}\sum_{a=1}^3\int''d\bm{x}\left[\nabla
 \left(\pi_a-\pi_a^0(\bm{m})\right)\right]^2+E^{\pi}_0(\bm{m}), \label{eq250}
\end{equation}
where $E_0(\bm{m})$ denotes the ground state energy. Clearly $\pi_a^0$ is the 
ground state of this Hamiltonian, and we can transform this 
Hamiltonian into the form of \eref{eq248}-\erefn{eq250} by integration by parts. 
Therefore, we regard \eref{eq250} as the simplified single-impurity
 Hamiltonian. As a result, the impurity part $H_{\rm imp}^{\pi}$ is derived as
\begin{align}
H_{\rm imp}^{\pi}\equiv&H^{\pi}-H_{\rm bulk}^{\pi}
                 =\frac{1}{2}J_{\pi}\sum_{a=1}^3\int''d\bm{x}\nabla\pi_a^0(\bm{m}) \nonumber\\
                &  \cdot\left[\nabla\pi_a^0(\bm{m})-2\nabla\pi_a\right]+E^{\pi}_0(\bm{m})\nonumber\\
               =&-J_{\pi}\sum_{a=1}^3\int''d\bm{x}\nabla\pi_a^0(\bm{m})\cdot\nabla\bm{\pi}+E_0^{\pi}(\bm{m}) \nonumber\\
                &+\frac{\pi J_{\pi}}{4}\left(\frac{r_0}{r^*}\right)^2(\mu_0^{\pi})^2(\bm{m}^4-m_1^4-m_2^4-m_3^4). \label{eq250.1}
\end{align}

\subsection{First Order Effect of the Interaction between the Impurity and 
       Magnetic Excitation}
Next we consider the effect of the coupling with magnetic excitation in the bulk region.
Including only the first order coupling of this kind, given by \eref{eq97}, 
the Hamiltonian for magnetic excitation becomes
\begin{align} 
H^{\psi}=&H_{\rm bulk}^{\psi}+H_{\rm imp}^{\psi},\nonumber\\
H_{\rm bulk}^{\psi}=&\frac{1}{2}J_{\psi}\sum_{a=1}^3\int'd\bm{x}
\left[(\nabla\psi_a)^2+{k_{\psi}}^2\psi_a^2\right], \label{eq152}
\end{align}
Just like the discussion above for nonmagnetic excitation, we consider only the 
dominant component of the magnetic excitation. The ground state configuration of 
$\bm{\psi}$ satisfies the Helmholtz 
 equation $\nabla^2\psi_a-{k_{\psi}}^2\psi_a=0$, 
and can be expanded in the polar coordinate as
\begin{equation}
\psi_a(r,\theta)=\sum_{m=0}^{\infty}c_m^{(a)}K_m(k_{\psi}r)
        \cos [m(\theta-\theta_{0,m}^{(a)})], \label{eq156}
\end{equation}
under the boundary condition 
$\psi_a(\bm{x})\rightarrow 0$ with $|\bm{x}|\rightarrow\infty$ 
\cite{Arfken}. Here $K_m(x)$ denotes the modified Bessel function of the second kind. 
We naturally expect that the angular momentum components $c_m^{(a)}$ with 
large $m$ are not dominant and neglect $c_m^{(a)}\quad(m>1)$. 
We call this "monopole-dipole approximation". We will justify 
this approximation by numerical calculations in \S 3.D. 
As a result, the ground state field configuration $\bm{\psi}$ is
\begin{equation}
\psi_a(r,\theta)=q_a^{\psi}K_0(k_{\psi}r)+K_1(k_{\psi}r)\frac{1}{r}\bm{r}
      \cdot\bm{\mu}_a^{\psi}, \label{eq160}
\end{equation}
where we introduced a charge $q_a^{\psi}$ and a dipole moment $\bm{\mu}_a^{\psi}$. 
Further, the bulk part of the energy becomes
\begin{equation}
E_{\rm bulk}^{\psi}=J_{\psi}\sum_a\left[\left(q_a^{\psi}\right)^2
{\cal E}_0(k_{\psi}r_0)+\left(\bm{\mu}_a^{\psi}\right)^2
  {\cal E}_1(k_{\psi}r_0)\right], \label{eq168} \\
\end{equation}
where we defined
\begin{align}
{\cal E}_0(x)=\pi xK_0(x)K_1(x) ,\nonumber\\
{\cal E}_1(x)=\frac{1}{2}\pi xK_0'(x)K_1'(x). \label{eq170}
\end{align}

Now we study the ground state of the total Hamiltonian \erefn{eq152}, using the 
monopole-dipole approximation. We deal with the ferromagnetic $(J<K/2)$ and 
antiferromagnetic $(J>K/2)$ regions separately, and first focus on the former. Here we 
focus the impurity located at "$E$" in \figref{unit_cell}, and 
show the results for the other cases in Appendix A.
In this case the energy of the lattice part 
Hamiltonian $H_{\rm imp}^{\psi}$ becomes
\begin{align}
E_{\rm imp}^{\psi}&=-\frac{\sqrt{3}}{2}J_{\psi}\sum_{a=1}^3\left[4\left
  \{K_0(k_{\psi} r_1)+K_1(k_{\psi}r_2)\right\}m_aq_a^{\psi}\right. \nonumber \\
& \left. +\frac{\sqrt{3}}{3}\left\{\frac{1}{r_1}K_1(k_{\psi}r_1)+
  \frac{4}{r_2}K_1(k_{\psi}r_2)\right\}m_a
              \hat{\bm{\kappa}}'_a\cdot\bm{\mu}_a^{\psi}\right], \label{eq172} 
\end{align}
where we defined
$\hat{\bm{\kappa}}'_1=(-\sqrt{3}/2,-1/2),\hat{\bm{\kappa}}'_2=(\sqrt{3}/2,\\-1/2),
\hat{\bm{\kappa}}'_3=(0,1)$, and 
the radial coordinates of shell sites $r_1=\sqrt{21}/3$ and $r_2=2\sqrt{3}/3$.
The total energy is the sum of the bulk part \erefn{eq168} and the impurity 
part \erefn{eq172}, and given by
\begin{align}
E&=J_{\psi}\sum_a\left[{\cal E}_0(k_{\psi} r_0)\left\{
 (q_a^{\psi}-q_0^{\psi}m_a)^2-(q_0^{\psi})^2m_a^2\right\}  \right.\nonumber \\
   &  \left.+{\cal E}_1(k_{\psi}r_0)\sum_{a}\left\{(\bm{\mu}_{a}^{\psi}-
             \mu_0^{\psi}m_a\hat{\bm{\kappa}}'_{a})^2-(\mu_0^{\psi})^2
               m_a^2\right\}\right], \label{eq179}
\end{align}
where we defined 
\begin{align}
q_0^{\psi}=&\frac{\sqrt{3}}{{\cal E}_0(k_{\psi}r_0)}\left(K_0(k_{\psi}r_1)
                  +K_0(k_{\psi} r_2)\right) \label{eq177} \\
\mu_0^{\psi}=&\frac{1}{4{\cal E}_1(k_{\psi}r_0)}\left
       (\frac{1}{r_1}K_1(k_{\psi}r_1)+\frac{4}{r_2}
                      K_1(k_{\psi}r_2)\right). \label{eq178}
\end{align}
Therefore, the charge and the dipole moment are determined by 
minimizing this energy as 
\begin{equation}
q_a^{\psi}=q_0^{\psi}m_a,\quad 
\bm{\mu}_{a}^{\psi}=\mu_0^{\psi}m_a\hat{\bm{\kappa}}'_{a}, \label{eq181}
\end{equation}
and the ground state energy is 
\begin{equation}
E_0^{\psi}(\bm{m})=-J_{\psi}\left[{\cal E}_0(k_{\psi}r_0)
\left(q_0^{\psi}\right)^2+{\cal E}_1(k_{\psi}r_0)\left(\mu_0^{\psi}\right)^2
                   \right] \label{eq182}
\end{equation}
Note that this ground state energy does not depend on the direction of the impurity magnetic 
moment $\bm{m}$. It means that magnetic excitation does not
 contribute to the anisotropy energy of the impurity magnetic moment, in the 
 first order coupling discussed here.

Next we turn to the antiferromagnetic region $(J>K/2)$. The lattice part of the energy 
becomes
\begin{align}
E_{\rm imp}^{\psi}=&-\frac{\sqrt{3}}{2}J_{\psi}\sum_{a=1}^3\left(
               \frac{1}{r_1}K_1(k_{\psi}r_1) \right. \nonumber\\
       &\left.+\frac{4}{r_2}K_1(k_{\psi}r_2)\right)m_a\hat{\bm{\kappa}}_a\cdot
                  \bm{\mu}_a^{\psi}. \label{eq183}
\end{align}
Combining this with the bulk part \erefn{eq170} yields the total energy
\begin{align}
E =&J_{\psi}\sum_a\left[{\cal E}_0(k_{\psi}r_0)\left(q_a^{\psi}\right)^2+
              {\cal E}_1(k_{\psi}r_0) \right.\nonumber \\
 &\left.
 \times\left\{(\bm{\mu}_a^{\psi}-\tilde{\mu}_0^{\psi}m_a\hat{\bm{\kappa}}_a)^2-
                \left(\tilde{\mu}_0^{\psi}\right)^2m_a^2
                  \right\}\right], \label{eq186}
\end{align}
where we defined $\tilde{\mu}_0^{\psi}=\sqrt{3}\mu_0^{\psi}$.
Therefore, the charge and the dipole moment are given by 
\begin{equation}
q_a^{\psi}=0,\quad \bm{\mu}_a^{\psi}=\tilde{\mu}_0^{\psi}m_a\hat{\bm{\kappa}}_a, 
                \label{eq188}
\end{equation}
and the ground state energy is
\begin{equation}
E_0^{\psi}(\bm{m})=-J_{\psi}{\cal E}_1(k_{\psi}r_0)
                   \left(\tilde{\mu}_0^{\psi}\right)^2. \label{eq189}
\end{equation}
Note that this energy does not depend on the direction of impurity magnetic moment 
$\bm{m}$, just as in the ferromagnetic region. 
It is notable that the first order effect of magnetic excitation in the bulk 
region does not yield the anisotropy of impurity magnetic moment 
in either ferromagnetic or antiferromagnetic region. We will see in the next 
subsection, however, magnetic excitation does contribute to the 
anisotropy energy through second order coupling.

 Just like in the previous subsection, we can modify the single-impurity 
 Hamiltonian \erefn{eq152} so that the monopole-dipole approximation is exact:
\begin{align}
 H^{\psi}=&\frac{1}{2}J_{\psi}\sum_{a=1}^3\int'' d\bm{x}\left\{\left[\nabla
          (\psi_a-\psi_a^0(\bm{m}))\right]^2 \right.\nonumber\\
  &   \left.   +k_{\psi}^2\left[\psi_a-\psi_a^0(\bm{m})\right]^2\right\}+E_0^{\psi}(\bm{m}), \label{eq276.5}
\end{align}
where $\psi_a^0$ denotes the field configuration of the ground state in 
monopole-dipole approximation. 
The impurity part $H_{\rm imp}^{\psi}$ is derived as
\begin{align}
 H^{\psi}_{\rm imp}=&\frac{1}{2}J_{\psi}\sum_{a=1}^3\int''d\bm{x}
    \left\{\nabla\psi_a^0\cdot[\nabla\psi_a^0-2\nabla\psi_a]\right. \nonumber\\
    &\left.\qquad+k_{\psi}^2\psi_a^0[\psi_a^0-2\psi_a]\right\}+E_0^{\psi}(\bm{m}) \nonumber\\
    =&-J_{\psi}\sum_{a=1}^3\int''d\bm{x}\left[\nabla\psi_a^0\cdot\nabla\psi_a+
        k_{\psi}^2\psi_a^0\psi_a\right]+E_0^{\psi}(\bm{m}) \nonumber\\
     &+J_{\psi}\sum_{a=1}^3\left[\left(q_a^{\psi}\right)^2{\cal E}_0(k_{\psi}r^*)
         +\left(\bm{\mu}_a^{\psi}\right){\cal E}_1(k_{\psi}r^*)\right]. \label{eq276.6}
\end{align}

Finally let us calculate the total induced magnetic moments 
 $\delta\bm{m}\equiv\sum_{i\notin {\rm core}}\langle \bm{S}\rangle$, and 
 the total squared induced magnetic moments $\delta m^2_{\rm sq}\equiv\sum_{i\notin {\rm core}}
 \langle \bm{S}\rangle^2$. 
In the ferromagentic region, these are derived by using magnetic excitation 
$\bm{\psi}$ as 
\begin{align}
\delta\bm{m}=&\frac{8}{3\sqrt{3}}\int''d\bm{r}\bm{\psi}=\frac{16\pi}{3\sqrt{3}}q_0^{\psi}\bm{m}\frac{r_0}{k_{\psi}}K_1(k_{\psi}r_0)  \label{ind1} \\
\delta m^2_{\rm sq}=&\frac{8}{3\sqrt{3}}\int''d\bm{r}\bm{\psi}^2  \nonumber\\  
                 =&\frac{8}{3\sqrt{3}}\pi r_0^2\left\{(q_0^{\psi})^2\left[K_1^2(k_{\psi}r_0)-K_0^2(k_{\psi}r_0)\right]\right. \nonumber\\
                   &\left.+\frac{1}{2}(\mu_0^{\psi})^2\left[K_0^2(k_{\psi}r_0)-K_1^2(k_{\psi}r_0)\right.\right.  \nonumber\\
                   &\left.\left.-\frac{2}{k_{\psi}r_0}K_0(k_{\psi}r_0)K_1(k_{\psi}r_0)
                    \right]\right\} \label{ind1.1}
\end {align}

These values diverge as $1/J$ for $J\rightarrow 0$. Note that this 
divergence should not be literally taken. 
For small $J$, magnetic excitation $\bm{\psi}$ on the sites near the 
impurity are not negligible and perturbative approach 
with regard to $\bm{\psi}$ breaks down. Therefore we anticipate actual 
divergence of induced magnetic moment $\delta\bm{m}$ is weaker 
than $1/J$. 
In the antiferromagnetic region, the present 
approach does not predict the value of the total induced magnetic 
moment $\delta\bm{m}$, since the field variable $\psi$ corresponds to the staggered 
magnetization. On the other hand, total squared moments $m_{\rm sq}$ can still be 
represented by the first line of \eref{ind1.1} and we obtain
\begin{align}
\delta m^2_{\rm sq} =&\frac{4}{3\sqrt{3}}\pi r_0^2(\tilde{\mu}_0^{\psi})^2\left[K_0^2(k_{\psi}r_0)-K_1^2(k_{\psi}r_0)  \right.\nonumber\\
                   &\left. -\frac{2}{k_{\psi}r_0}K_0(k_{\psi}r_0)K_1(k_{\psi}r_0)\right], \label{ind1.4}
\end{align}
which diverges as $1/(1-K/J)$ for $J\rightarrow K$. Again, actual 
divergence is expected to be weaker. 
We will numerically investigate these two representations of induced magnetic moment in \S 4.6.

\subsection{Second Order Effect of the Interaction between the Impurity and 
                 Bulk Excitation}
In the previous argument, we showed that the magnetic anisotropy emerges 
from the first order coupling to nonmagnetic excitation in the 
bulk region, and that the anisotropy is corner-cubic one represented by \eref{eq137}. 
We will confirm this, by means of numerical calculations in the next subsection, 
but we will also find different types of anisotropy 
appear when $J/K$ is small. They originate from  higher-order effects of the 
coupling between impurity magnetic moment and bulk excitations. 
In this subsection we examine the effects of the second order coupling and investigate
 the magnetic anisotropy. 
 Adding the three second-order terms 
  \erefn{eq94}-\erefn{eq99}, 
the total Hamiltonian is now given by
\begin{align}
H^{(2)}=&H_{\rm bulk}^{\pi}+H_{\rm bulk}^{\psi}+H_{\rm imp}^{\pi}+
                                        H_{\rm imp}^{\psi} \nonumber\\
 &+H_{\rm imp}^{\pi^2}+H_{\rm imp}^{\psi^2}+H_{\rm imp}^{\pi\psi} \label{eq194.1}
\end{align}
where the bulk parts $H_{\rm bulk}^{\pi}$ and $H_{\rm bulk}^{\psi}$ are given 
in \eref{eq108} and \erefn{eq152}, respectively.
As for the bulk region, we have employed the dipole approximation for the 
nonmagnetic excitation $\bm{\pi}$, and the monopole-dipole approximation 
for the magnetic excitation $\bm{\psi}$. Here we introduce two 
additional approximations, in order to  simplify the problem. First, we 
consider only the lowest-order nonzero angular momentum component for the 
magnetic excitation $\bm{\psi}$. As is clear from the previous results 
\erefn{eq181} and \erefn{eq188}, it corresponds to the monopole approximation 
for the ferromagnetic region $(J<K/2)$ and dipole approximation for the antiferromagnetic 
region $(J>K/2)$. Numerical calculations in the next subsection shows that 
these lowest-order  moments are more than several 
times larger than the higher order moments.
 Second, we assume that the dipole moments for the field 
variables point to the direction obtained in the first order calculation. Again, numerical 
calculations will verify this approximation. As a result of these two 
approximations, the field variables are represented as
\begin{align}
\pi_a(\bm{r})=&\frac{r_0}{r^2}\mu_a^{\pi}\hat{\bm{\kappa}}_a\cdot\bm{r},
                                     \label{eq196} \\
\psi_a(\bm{r})=&q_a^{\psi}K_0(k_{\psi}r)\ &F.M. \label{eq197} \\
\psi_a(\bm{r})=&K_1(k_{\psi}r)\frac{1}{r}\mu_a^{\pi}
             \cdot\hat{\bm{\kappa}}_a\cdot\bm{r} &A.F.M., \label{eq199}
\end{align}
with scalar variables $\mu_a^{\pi},q_a^{\psi},$ and $\mu_a^{\psi}$. 
Note that it is straightforward, though not shown here, 
to extend the following argument to more general situation that we do not employ 
these approximations.

Now let us calculate the ground state energy of the Hamiltonian \erefn{eq194.1} 
within the approximations above, and we first focus on the 
ferromagnetic region $(J<K/2)$. For the configuration \erefn{eq196}-\erefn{eq199}, 
the energy becomes
\begin{equation}
E^{(2)}=\sum_{I,J=\pi,\psi}\bm{g}_I\cdot\hat{M}_{IJ}\bm{g}_J-2
                   \sum_{I=\pi,\psi}\bm{f}_I\cdot\bm{g}_I \label{eq209}
\end{equation}
where $\bm{f}_I$ and $\bm{g}_I,\ (I=\pi,\psi)$ are three-component vectors
\begin{align}
\bm{f}_{\pi}=&c_{\pi}\mu_0^{\pi}\left(m_2m_3,m_3m_1,m_1m_2\right)^T \label{eq211}\\
\bm{f}_{\psi}=&c_{\psi}q_0^{\psi}\left(m_1,m_2,m_3\right)^T \label{eq211.5} \\
\bm{g}_{\pi}=&\left(\mu_1^{\pi},\mu_2^{\pi},\mu_3^{\pi}\right)^T \label{eq210} \\
\bm{g}_{\psi}=&\left(q_1^{\psi},q_2^{\psi},q_3^{\psi}\right)^T, \label{eq210.5} 
\end{align}
$\hat{M}_{IJ}$'s are three-by-three matrices
\begin{align}
\left(M_{\pi\pi}\right)_{ij}&=c_{\pi}\delta_{ij}+\frac{1}{2}c_{\pi^2}(1-\delta_{ij})m_im_j, \nonumber\\
\left(M_{\psi\psi}\right)_{ij}&=c_{\psi}\delta_{ij}+\frac{1}{2}c_{\psi^2}(1-\delta{ij})m_im_j, \nonumber\\
\left(M_{\pi\psi}\right)_{ij}&=\left(M_{\psi\pi}\right)_{ij}=\frac{1}{4}c_{\pi\psi}\sum_{k=1}^3|\epsilon_{ijk}|m_k, \label{eq212}
\end{align}
with coefficients
\begin{align}
c_{\pi}=&\frac{\pi J_{\pi}}{2} r_0^2,\  
c_{\psi}=J_{\psi}{\cal E}_0\left(k^\psi r_0\right),\  
c_{\pi^2}=-\frac{369}{98}Kr_0^2,\label{eq204.5} \\
c_{\psi^2}=&2K\left\{K_0\left(k_{\psi}r_1\right)^2+K_0\left(k_{\psi}r_2\right)^2
                                  \right\} \label{eq201} \\
c_{\pi\psi}=&-2(2J-K)r_0\left\{\frac{1}{2r_1^2}K_0\left(k_{\psi}r_1\right)+
          \frac{2}{r_2^2}K_0\left(k_{\psi}r_2\right)\right\} \label{eq206}
\end{align}
The ground state energy of the Hamiltonian \erefn{eq209} is readily derived as
\begin{equation}
E_0(\bm{m})=-\sum_{I,J=\pi,\psi}\bm{f}_I
            \cdot(\hat{M}_{IJ}-\hat{\Sigma}_{IJ})^{-1}\bm{f}_J \label{eq215}
\end{equation}
Here, the self-energy part is given as
\begin{align}
  \hat{\Sigma}_{\pi\pi}=&\hat{M}_{\pi\psi}
                           \hat{M}_{\psi\psi}^{-1}\hat{M}_{\psi\pi},\  
  &\hat{\Sigma}_{\psi\psi}=\hat{M}_{\psi\pi}\hat{M}_{\pi\pi}^{-1}
                                     \hat{M}_{\pi\psi} \label{eq215.1} \\
  \hat{\Sigma}_{\pi\psi}=&\hat{M}_{\psi\psi}
                       \hat{M}_{\psi\pi}^{-1}\hat{M}_{\pi\pi},\ 
  &\hat{\Sigma}_{\psi\pi}=\hat{M}_{\pi\pi}
                   \hat{M}_{\pi\psi}^{-1}\hat{M}_{\psi\psi} \label{eq215.2}
\end{align}
and represents the effects of coupling between nonmagnetic and magnetic 
excitations. (We do not show the explicit form of the ground state energy, 
since it is quite lengthy.) Later we investigate the dependency of the 
ground state energy on the coupling constants $J/K$.

Next, we turn to the antiferromagnetic region $(J>K/2)$. The ground state energy 
is similarly calculated and the result is given by 
replacing the variables in eqs. \erefn{eq215}-\erefn{eq215.2} 
by those for the antiferromagnetic region as 
\begin{align}
c_{{\psi}^2}=&-K\left[\left(\frac{1}{r_1}K_1(k_{\psi})\right)^2-2
        \left(\frac{1}{r_2}K_1(k_{\psi}r_2)\right)^2\right] \label{eq203} \\
c_{\pi{\psi}}=&2(2J-K)r_0\left[-\frac{13}{4r_1^3}K_1(k_{\psi}r_1)+
                    \frac{2}{r_2^3}K_1(k_{\psi}r_2)\right] \label{eq208} \\
c_{{\psi}}=&J_{\psi}{\cal E}_1(k_{\psi}r_0). \label{eq243}
\end{align}
and $q_0^{\psi}$ in \eref{eq211.5} and $q_a^{\psi}\ (a=1,2,3)$ in \eref{eq210.5} 
should be replaced with $\tilde{\mu}_0^{\psi}$ and $\mu_a^{\psi}$ ,respectively.

We show later in \figref{sclen} (b) the ground state energy \erefn{eq215} 
as a function of $J/K$ when the impurity magnetic moment is fixed along the representative 
directions.
  We can see that the energy now depends on $J/K$, unlike the results with 
the first order coupling \erefn{eq137} and \erefn{eq182}. This is an effect of the 
  second-order couplings $c_{\pi^2},\ c_{\psi^2}$ and $c_{\pi\psi}$. 
In the parameter region $0<J/K<0.077$,
(100) direction is an easy direction, i.e. the anisotropy is "face-cubic". 
We will compare this result with numerical results in the next subsection.

\subsection{Numerical Calculation}
We have shown analytically the appearance of the magnetic anisotropy and the 
configuration of the ground state, based on the effective field theory for 
the low-energy excitations in the bulk region. In this subsection we numerically 
solve this one impurity problem, in order to compare with the approximate 
analytical results in the previous subsections.

\begin{figure}[b]
\centering
\includegraphics[scale=0.44,viewport=0 0 276 257]{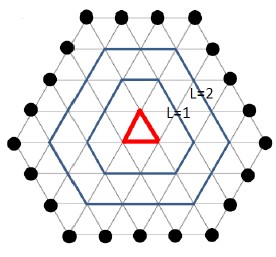}
\caption{The geometry of the finite size sample used in the numerical 
calculation. The case of size $L=2$ is shown. States on the sites denoted by black 
circles are fixed to the mean-field ground state \erefn{mfgs}. }
\label{lattice_1body}
\end{figure}

We calculated the ground-state energy and spin configuration, by 
numerical calculations with finite-size clusters up to $L=40$, including 5292 sites. The geometry of 
the cluster is depicted in \figref{lattice_1body}. 
$L$ labels the layer of sites away from the impurity bond triad placed at the origin, 
and it characterizes the cluster size. 
To minimize finite-size effects, we set the spin wavefunctions on the outer-boundary 
sites (shown by black circles in \figref{lattice_1body})
to the bulk values \erefn{mfgs}. With this boundary condition, we minimize the 
energy \erefn{eq9} by optimizing wavefunctions of each spin. 
As for the ferromagnetic coupling between core sites $J'<0$, we examined both 
of finite and infinite case. 
For the infinite case $(J'=-\infty)$, we fixed the direction of impurity magnetic moment and examined the 
ground state for each value of the impurity magnetic moment, in order to 
compare with the analytical results.

\begin{figure}[!b]
\centering
\includegraphics[scale=0.44,viewport=0 0 522 773]{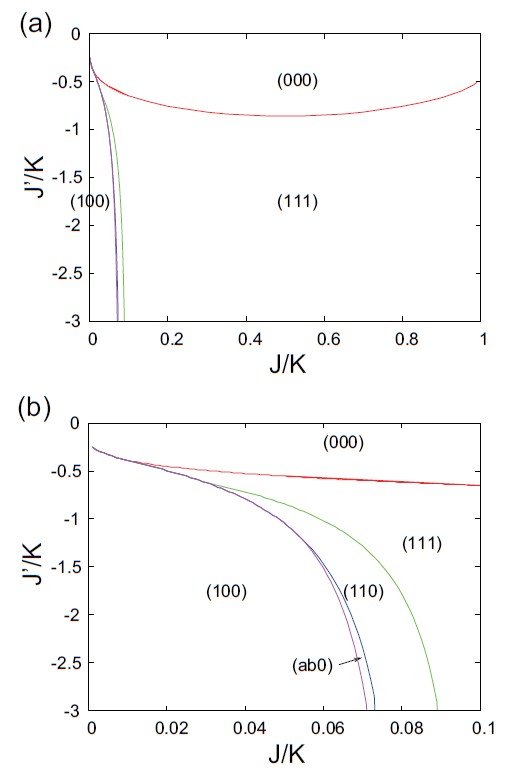}
\caption{Phase diagram of the ground state in $J$-$J'$ parameter space. 
(a) $0\leq J/K \leq 1$, (b) $0 \leq J/K\leq 0.1$.}
\label{phase}
\end{figure}

\begin{table}[b]
    \begin{center}
     \caption{Labeling of five phases and easy directions of impurity magnetic moment
           in each phase.}
    \begin{tabular}{|c|c|}
      \hline
      label  & easy directions \\ \hline\hline
      (000)  & $(0,0,0)$  \\ \hline 
      (111)  & $\frac{1}{\sqrt{3}}(\pm 1,\pm 1,\pm 1)$ \\ \hline 
      (110)  & $\frac{1}{\sqrt{2}}(\pm 1,\pm 1,0),\ \frac{1}{\sqrt{2}}
                (\pm 1,0,\pm 1),\ \frac{1}{\sqrt{2}}(0,\pm 1,\pm 1)$ \\ \hline
      (ab0)  & $\displaystyle (\pm a,\pm b,0),\ (\pm b,\pm a,0),
         \ (\pm a,0,\pm b),\atop \displaystyle(\pm b,0,\pm a),
         \ (0,\pm a,\pm b),\ (0,\pm b,\pm a)\ (a^2+b^2=1)$ \\ \hline
      (100)  & $(\pm 1,0,0),\ (0,\pm 1,0),\ (0,0,\pm 1)$ \\ \hline
   \end{tabular}
   \label{phases}
\end{center}
\end{table}
First let us show the results for finite $J'$. \figref{phase} shows the phase diagram 
and each phase is characterized by the easy directions of the average
 magnetic moment of the three core sites: $\overline{\bm{m}}\equiv\frac{1}{3}
 \sum_{i\in{\rm core}}\bm{m}_i$. We find that there are five phases with 
 different easy axes and label these phases as shown in \tabref{phases}.
The $(111)$ phase has a corner-cubic anisotropy, 
and covers the largest region of the coupling $J/K$, when $|J'|$ is sufficiently 
large. On the other hand, the $(100)$ phase has 
a face-cubic anisotropy and appears near the $J=0$ line. 
Moreover, between these two phases we find the $(110)$ phase, which has an 
edge-cubic anisotropy, and the least symmetric $(ab0)$ phase.

\begin{figure}[!t]
\centering
\includegraphics[scale=0.44,viewport=0 0 444 769]{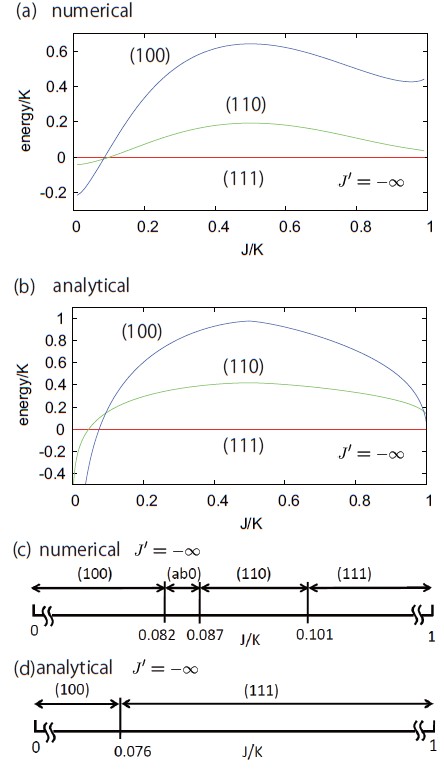}
\caption{(a)(b)Energies of the (111) and (110) states measured 
      from the energy of the (100) state in the limit $J'=-\infty$.
(a):numerical result, (b):analytical result.
(c)(d) The ground state phase diagram in the limit $J'=-\infty$. (c):numerical result, 
       (d):analytical result.}
\label{sclen}
\end{figure}
Second we show the results in the limit $J'=-\infty$. \figref{sclen}(a) and (c) show the energy 
difference between three representative phases and the phase diagram, respectively. 
 The corresponding analytical results are shown in 
 \figref{sclen}(b) and (d). The four phases observed at finite $|J'|$, survive 
 in this $J'=-\infty$ limit, as shown in \figref{sclen} (c). It implies that
 the finiteness of $|J'|$ is not essential for the presence of these anisotropies 
 of the impurity magnetic moment. Further, the appearance of the two phases 
 (100) and (111), which cover large regions in the phase diagram, qualitatively 
 agrees with the analytical results shown in \figref{sclen} (d), whereas 
the other two phases appear only in numerical calculation. 
The dependency of the magnetic anisotropy energy, shown in 
\figref{sclen}(a), qualitatively agrees with the analytical results in 
\figref{sclen}(b), although the difference becomes significant near 
$J/K=0$ or $1$. This discrepancy can possibly be ascribed to the breakdown of 
the perturbative treatment of the magnetic excitation. Near the onset of 
 magnetically ordered phases ($J/K=0$ or $1$), the system becomes sensitive 
 to the magnetic impurity and the amplitude of the 
 field for magnetic excitation increases.

\begin{figure}[t]
\centering
\includegraphics[scale=0.44,viewport=0 0 549 547]{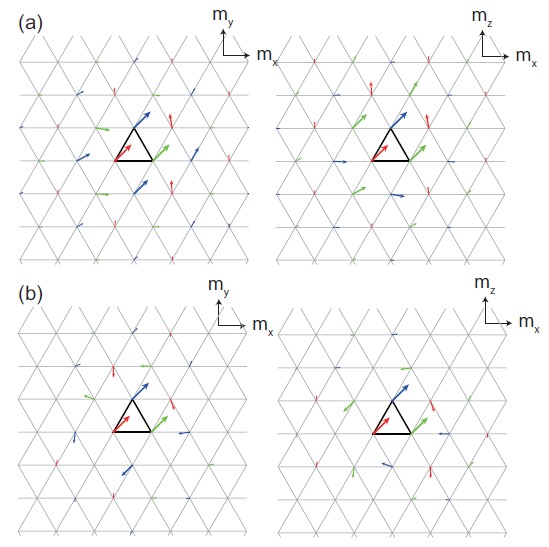}
\caption{Numerical result of spatial distribution of induced magnetic moments in the limit $J'=-\infty$ and 
    $\bm{m}=\frac{1}{\sqrt{3}}(1,1,1)$. (a) ferromagnetic case $(J/K=0.05)$, 
    (b) antiferromagnetic case $(J/K=0.95)$. Projections to XY and XZ planes are 
    shown in the left and right panel, respectively.}
\label{magn}
\end{figure}

\begin{figure}
\centering
\includegraphics[scale=0.44,viewport=0 0 573 1137]{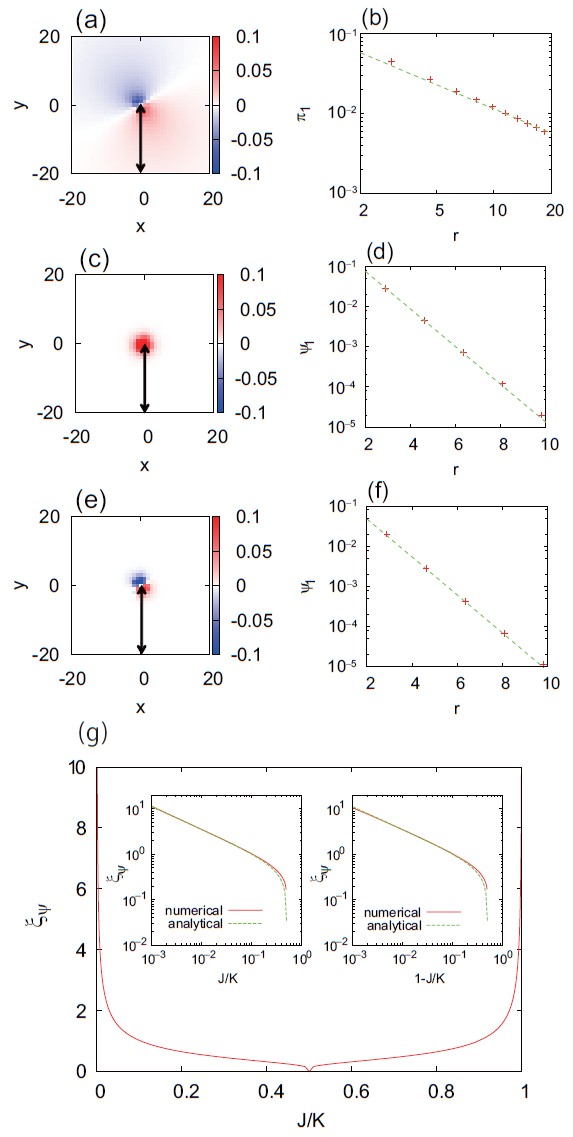}
\caption{Numerical results of spatial distribution of the field variables for $J'=-\infty$ and 
    $\bm{m}=\frac{1}{\sqrt{3}}(1,1,1)$. 
(a):Nonmagnetic excitation $\pi_1$ for J/K=0.5. (b):Log-log plot of $\pi_1$ 
         along the arrow in (a).
(c):Magnetic excitation $\psi_1$ for J/K=0.1. (d):Semilog plot of $\psi_1$ 
          along the arrow in (c).
(e):$\psi_1$ for J/K=0.9. (f):Semilog plot of $\psi_1$ along the arrow in (e). 
(g):Decay length of $\psi_1$. Two insets show log-log plots of $\xi_{\psi}=1/k_{\psi}$ near $J/K=0$ and $1$, 
   together with analytical prediction \erefn{eq84}. }

\label{field}
\end{figure}

Then we examine the ground state spin configuration $\langle \bm{S}_i\rangle$ in the limit $J'=-\infty$. 
As a typical example, we fix the impurity magnetic moment 
as $\bm{m}=\frac{1}{\sqrt{3}}(1,1,1)$. 
Before detailed argument, 
we show induced magnetic moments around the impurity in the ferromagnetic region $(J/K=0.05)$ 
in \figref{magn} (a), and in the antiferromagnetic region $(J/K=0.95)$ in (b). 
In both regions, induced magnetic moments form a complex noncoplanar pattern. 
This noncoplanarity can be understood as the result of antiferro quadrupolar order. 
As mentioned in \S 3.2, magnetic moments are perpendicular to the directors, which are orthogonal 
between different sublattices. Note that this orthogonality is violated around the impurity, 
due to the nonmagnetic excitation. 

 In order to compare the results with analytical ones, we converted the spin 
 configuration to the field variables for nonmagnetic and magnetic excitation, 
 by solving equations \erefn{eq88} and \erefn{eq89}.
The results are shown in \figref{field}.
Nonmagnetic excitation $\pi_1$ around the impurity is plotted for two-dimensional 
space in \figref{field}(a) and also along one direction 
in \figref{field}(b). 
We can see that it shows dipole-like angle dependence, changing sign once around 
the impurity site, and its radial dependence decays as $1/r$. This result 
verifies the dipole approximation employed in 
\S 4.2 at least on the qualitative level. The direction of the principal 
axis is $\hat{\bm{\kappa}}_1=(1/2,-\sqrt{3}/2)$ and coincides with that in the 
analytical result \erefn{eq134}. Next we turn to the magnetic excitation 
 $\bm{\psi}$, and show it in \figref{field} (c) and (d) for 
the ferromagnetic region ($J/K=0.1$) and \figref{field} (e)(f) for the 
antiferromagnetic region ($J/K=0.9$). Its angular part has isotropic (dipole-like) 
in the ferromagnetic (antiferromagnetic) region, and the radial dependence 
decays exponentially in both cases. These results 
verify the monopole-dipole approximation we used in \S 4.1. Again, for the 
antiferromagnetic region, the direction of the principal axis
 is $\hat{\bm{\kappa}}_1$ and agrees with that in the analytical 
 result \erefn{eq188}. In \figref{field}(g) we present the decay length $\xi_{\psi}$ of magnetic excitation $\bm{\psi}$, 
 which is derived by fitting the radial dependence of $\bm{\psi}$ to the exponential form $\psi(r)\propto \exp(-r/\xi_{\psi})$. 
In the regions near the phase boundary with ferromagnetic (antiferromagnetic) phase, $0<J/K<0.1$ $(0.9<J/K<1)$, 
$\xi_{\psi}$ agrees well with its analytical value $\xi_{\psi}=1/k_{\psi}$. In the intermediate region $0.1<J/K<0.9$ with 
$\xi_{\psi}\lesssim 1$, on the 
other hand, there exist discrepancy between numerical and analytical values. This discrepancy is ascribed to 
the coarse graining of the lattice model, which is justified only when the characteristic length scale $\xi_{\psi}$ of the 
theory is much larger than the lattice constant. 

\begin{table}[b]
    \begin{center}
      
     \caption{Angular momentum components of the field variables for 
        $\bm{m}=\frac{1}{\sqrt{3}}(1,1,1)$.
    $q, \mu,$ and $d$ denote the absolute values of monopole, dipole, and 
            quadrupole component respectively.}
    \begin{tabular}{|c|c|c|c|c|}\cline{3-5}
    \multicolumn{2}{c|}{}                                              & \multicolumn{3}{|c|}{$J/K$}     \\ \cline{3-5}
    \multicolumn{2}{c|}{}                                              & 0.1      & 0.5     &   0.9      \\ \hline\hline
    \multirow{2}{*}{$\pi$}                & $\mu^{\pi}$               & 0.086    &  0.138  &  0.125       \\ \cline{2-5}
                                          & $d^{\pi}$                 & 0.013    &  0.003  &  0.029       \\ \hline\hline
    \multirow{3}{*}{$\psi$}               & $q^{\psi}$                & 0.529    &  -      &  0.000       \\ \cline{2-5}
                                          & $\mu^{\psi}$              & 0.113    &  -      &  0.925       \\ \cline{2-5}
                                          & $d^{\psi}$                & 0.028    &  -      &  0.207       \\ \hline
    \end{tabular}
    \label{num_moments}
    \end{center}
   \end{table}
\begin{figure}[t]
\centering
\includegraphics[scale=0.44,viewport=0 0 561 665]{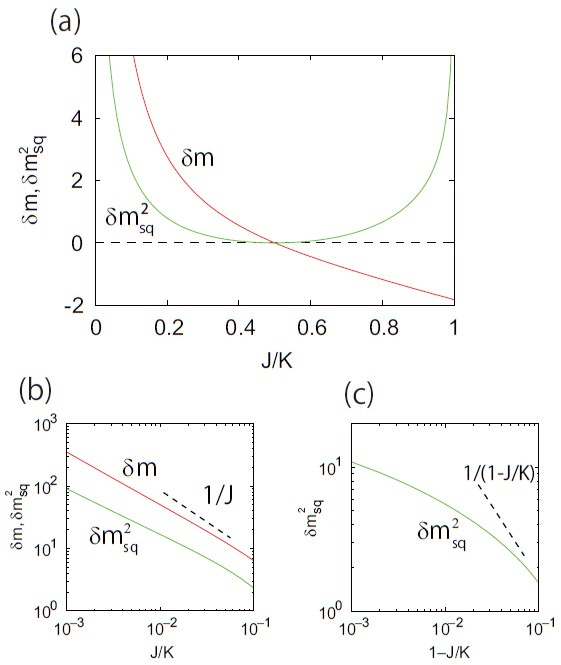}
\caption{(a): Induced magnetic moment $\delta\bm{m}$ measured along the direction of impurity 
magnetic moment $\delta m=\bm{m}\cdot\delta\bm{m}$ and total squared induced moment $\delta m^2_{\rm sq}$, 
in the limit $J'=-\infty$ and $\bm{m}=\frac{1}{\sqrt{3}}(1,1,1)$. 
(b): Log-log plot in the ferromagnetic region. 
(c): Log-log plot in the antiferromagnetic region. Notice the horizontal axis is $1-J/K$.  
Cluster size is carefully chosen 
so that the finite siza effect is negligible (less than 1\%).}
\label{induced_moment}
\end{figure}

In order to further examine the validity of the simplification of the angular 
part for $\bm{\pi}$ and $\bm{\psi}$, we calculated the amplitude of each partial wave 
in eqs. \erefn{eq111} and \erefn{eq156} up to quadrupole ($m=2$) component.
The results are listed in \tabref{num_moments}. We can see that the dominant 
component (i.e. ($\mu^{\pi}(J=0.1,0.5,0.9),\ 
q^{\psi}(J=0.1),\ \mu^{\psi}(J=0.9)$) is more than several times larger than 
the other components. This justifies our approximation 
that only single dominant component is taken into account.

Finally let us examine induced magnetic moments outside the impurity 
$\delta\bm{m}$. \figref{induced_moment} shows total induced magnetic moment $\delta\bm{m}$ and 
total squared induced magnetic moment $\delta m^2_{\rm sq}$, defined in \S 4.3, as 
functions of $J/K$ for (111) phase. In the ferromagnetic region, $\delta\bm{m}$ 
is parallel to the impurity magnetic moment and $\delta\bm{m}$ and $\delta m^2_{\rm sq}$ diverge with $J\rightarrow 0$. 
This divergence is slightly weaker than $1/J$, as mentioned in \S 4.3. 
In the antiferromagnetic region, 
is antiparallel to the impurity magnetic moment, as anticipated from 
antiferromagnetic interaction. 
Contrary to the ferromagnetic region, 
the absolute value of induced magnetic moment remains finite as $J\rightarrow K$.
The squared moment $m_{\rm sq}$ diverges with $J/K\rightarrow 1$ and the divergence is 
weaker than analytically derived behavior, $(1-J/K)^{-1}$.

\section{Interaction between Impurities}

In the previous section, we showed that magnetic and nonmagnetic excitations 
are induced around an impurity magnetic moment. 
When there exist multiple impurities, 
interference of these excitations causes interaction between impurities. 
In this section, we study this impurity-impurity
 interaction, using both analytical and numerical methods. We will show that 
 nonmagnetic excitation yields long-range 
 interaction with spatial anisotropy.
 
In order to derive interaction between impurities, let us consider a system 
with two impurities with their magnetic moments fixed, and calculate its 
ground state energy. As in the previous section, we  describe the bulk part by 
 the continuum effective model, as shown in \figref{2body_approx}. Regarding 
 the impurity part $H_{\rm imp}$, we consider only the first order coupling of 
 impurity magnetic moment and bulk excitations. As a result we can decompose 
 the Hamiltonian in the magnetic and nonmagnetic parts as 
 $H=H^{\pi}(\{\bm{\pi}\})+H^{\psi}(\{\bm{\psi}\})$, and study the ground state 
 of each part separately, just like the discussion in the previous section.

 For each single
 impurity, we take only the most dominant angular component of excitations,
  $\bm{\pi}$ and $\bm{\psi}$, 
 in the bulk region. It corresponds to the dipole approximation for 
 the nonmagnetic excitation $\bm{\pi}$. As for the magnetic excitation $\bm{\psi}$, 
 the dominant mode is a dipole component in the antiferromagnetic region $(J/K>2)$, 
 and a scalar component in the ferromagnetic region $(J<K/2)$. We use the minimal 
 models of impurity \erefn{eq250} and \erefn{eq276.5}, for which these 
 approximations are exact. In these models, we take the limit $r^*\rightarrow 0$, 
 where $r^*$ denotes the  radius of the void of the bulk region shown in 
 \figref{2body_approx}(b), since finiteness of $r^*$ is irrelevant to the 
 asymptotic behavior of the  interaction energy as the separation tends to 
 infinity.
 
 We will analytically derive the interaction mediated by nonmagnetic and magnetic 
 excitations in \S 5.1 and \S 5.2, respectively, 
and then proceed to numerical analysis in \S 5.3.

\subsection{Interaction Mediated by Nonmagnetic Excitation}
Let us derive the interaction energy between two impurities mediated by 
nonmagnetic excitation in the bulk. Using the simplified one-impurity 
Hamiltonian \erefn{eq248}, \erefn{eq249} and \erefn{eq250.1}, the total 
Hamiltonian with two impurities is given by
\begin{align}
H^{\pi}=&H^{\pi}_{\rm bulk}+H^{\pi(a)}_{\rm imp}+H^{\pi(b)}_{\rm imp} \nonumber\\
=&\frac{1}{2}J_{\pi}\int d\bm{x}\left\{\nabla\left(\bm{\pi}-\left(\bm{\pi}^{0(\alpha)}+
                     \bm{\pi}^{0(\beta)}\right)\right)\right\}^2 \nonumber \\
 &               -J_{\pi}\int d\bm{x}\nabla\bm{\pi}^{0(\alpha)}
                                  \cdot\nabla\bm{\pi}^{0(\beta)}  \nonumber\\
 &               +E_0(\bm{m}^{(\alpha)})+E_0(\bm{m}^{(\beta)}), \label{eq259}
\end{align}
where $\bm{m}^{(i)}$ and $\bm{\pi}^{0(i)},\ (i=\alpha,\beta)$ denote the impurity 
magnetic moments and the ground state configuration in the presence of 
the individual impurity, respectively. 
$i=\alpha,\beta$ denotes the impurity index. Note that $\bm{\pi}^0$ itself 
depends on the magnetic moment at the impurity, 
$\bm{\pi}^{0(i)}=\bm{\pi}^{0}(\bm{x};\bm{m}^{(i)})$. 
We can define the interaction energy 
$H_{\rm int}^{\pi}$ as the difference between the ground state energy 
of the whole system, which is calculated for \eref{eq259}, and the sum of 
the two ground state energies of the one-impurity Hamiltonians:
\begin{align}
H_{\rm int}^{\pi}\equiv& E_{\rm g.s.}-\left(E_0(\bm{m}^{(\alpha)})
                             +E_0(\bm{m}^{(\beta)})\right) \nonumber\\
       =&-J_{\pi}\int d^2 x\nabla\bm{\pi}^{0(\alpha)}
                            \cdot\nabla\bm{\pi}^{0(\beta)} \label{eq261}
\end{align}
Using the expression of the one-impurity ground state \erefn{eq114} and 
                           \erefn{eq134}, the interaction energy is expressed 
in terms of three-by-three matrices 
\begin{align}
H_{\rm int}^{\pi}=&\pi J_{\pi}\frac{r_0^2(\mu_0^{\pi})^2}{r_{\alpha\beta}^2}
              {\rm Tr}\left(T^{(\alpha)}fT^{(\beta)}\right) \label{eq262} \\
\left(T^{(x)}\right)_{ab}=&m_a^{(x)}m_b^{(x)}-\delta_{ab}{m_{a}^{(x)}}^2,
                      \quad (x=\alpha,\beta) \label{eq263} \\
f=&\left(\begin{array}{ccc}
\cos(2\theta+\frac{2\pi}{3}) &0 &0\\
0 & \cos(2\theta-\frac{2\pi}{3}) & 0 \\
0 & 0 & \cos 2\theta
\end{array}\right), \label{eq264}
\end{align}
where the separation between the two impurities is expressed as 
$\bm{r}_{\beta}-\bm{r}_{\alpha} \equiv\bm{r}_{\alpha\beta}=
r_{\alpha\beta}(\cos\theta, \sin\theta)$.
The interaction is biquadratic with regard to impurity magnetic moments 
$\bm{m}$, 
and its radial part has $r^{-2}$ dependence, whereas its angular part has 
dipole-dipole like anisotropy. Note that nonzero elements of $T$ matrix are 
nothing but $t_{2g}$ part of spin quadrupole moment defined by \eref{q2}. 
Therefore the interaction \erefn{eq262} can also  be interpreted as 
interaction between quadrupole moments of impurity spins.

\subsection{Interaction Mediated by Magnetic Excitation}
Second let us derive the interaction between impurities mediated by magnetic 
excitations, using simplified one-impurity Hamiltonian \erefn{eq276.5}.
The result is different between the ferromagnetic and antiferromagnetic 
regions $(J\gtrless K/2)$.

In the ferromagnetic region $(J<K/2)$, the interaction energy is
\begin{equation}
H_{\rm int}^{\psi}=-2\pi J_{\psi}(q_0^{\psi})^2K_0(k_{\psi}r_{\alpha\beta})
      \bm{m}^{(\alpha)}\cdot\bm{m}^{(\beta)}, \label{eq288}
\end{equation}
where $K_0(x)$ denotes the modified Bessel function of the second kind. 
The long-range asymptotic form is given by
\begin{equation}
H_{\rm int}^{\psi}\sim-\sqrt{\frac{2\pi^3}{k_{\psi}r_{\alpha\beta}}}J_{\psi}
(q_0^{\psi})^2e^{-k_{\psi}r_{\alpha\beta}}\bm{m}^{(\alpha)}\cdot\bm{m}^{(\beta)}, 
                                             \label{eq314.2}
\end{equation}
where we used the asymptotic form of the modified Bessel function of the second 
 kind $K_{\nu}(z)$ as $z\rightarrow\infty$:
\begin{equation}
K_{\nu}(z)=\sqrt{\frac{\pi}{2z}}e^{-z}\left[1+\frac{4\nu^2-1}{8z}+\cdots\right] 
                 \label{eq314.1}
\end{equation}
The expression \erefn{eq314.2} shows that the interaction is ferromagnetic and 
isotropic in both of the spin space and the real space,
and exponentially decays with the characteristic length scale of magnetic 
excitation, 
\begin{equation}
\frac{1}{k_{\psi}}=\frac{1}{2}\left(\frac{K}{|K-2J|}-1\right)^{-1/2}. \label{eq314.3}
\end{equation}

In the antiferromagnetic region $(J>K/2)$ , the interaction energy is 
\begin{equation}
H_{\rm int}^{\psi}=\pi J_{\psi}(\tilde{\mu}_0^{\psi})^2\bm{m}^{(\alpha)}\cdot
  \left[K_2(k_{\psi}r_{\alpha\beta})f+K_0(k_{\psi}r_{\alpha\beta})\bm{1}\right]
\bm{m}^{(\beta)}, \label{eq298.6}
\end{equation}
where the form factor $f$ is defined in \eref{eq264} 
Its long-range asymptotic form is
\begin{equation}
H_{\rm int}^{\psi}=\sqrt{\frac{\pi^3}{2k_{\psi}r_{\alpha\beta}}}
       J_{\psi}(\mu_0^{\psi})^2 e^{-k_{\psi}r_{\alpha\beta}}
\bm{m}^{(\alpha)}\cdot\left[f-\bm{1}\right]\bm{m}^{(\beta)}. \label{eq298.7}
\end{equation}
This result shows that, the bilinear coupling of the magnetic
 moments and the exponential decay of the radial part are common with the 
 ferromagnetic region, but the interaction has anisotropy. 
 Its angular part contains dipole-dipole like anisotropic terms inadition to 
 isotropic and ferromagnetic term.

\subsection{Numerical Calculation}
We have analytically derived the interaction between impurity magnetic moments, 
making use of the continuum field theory of the low energy excitation
 in the bulk region and simplified coupling between each impurity and the 
 bulk. Now we numerically calculate interaction 
 energy, in order to check the validity of these simplifications. 

We calculated the ground state energy of a finite-size system with two 
impurities of $L=40$ with the fixed boundary condition as explained in \S 4.2. 
Here $L$ denotes the number of layers around the origin and impurities are located 
symmetrically around the origin. 
Then we determined the interaction energy by subtracting from the ground-state 
energy one-impurity contributions, 
which were also derived numerically. We took the limit $J'=-\infty$ and fixed 
the impurity magnetic moments, 
for proper comparison with the analytical results in the previous subsection. 
In addition, we consider the cases
$(\bm{m}^{(\alpha)},\bm{m}^{(\beta)})=(\bm{m}_A,\bm{m}_B),\ (\bm{m}_A,-\bm{m}_B)$, 
where $\bm{m}_A\equiv\frac{1}{\sqrt{3}}
(1,1,1),\ \bm{m}_B\equiv\frac{1}{\sqrt{3}}(1,-1,-1)$. 

\begin{figure}
\centering
\includegraphics[scale=0.44,viewport=0 0 526 720]{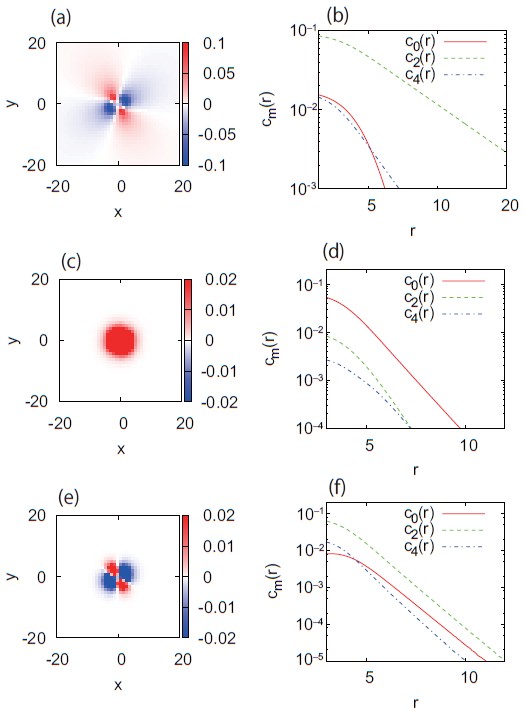}
\caption{Distance dependence of (a) nonmagnetic interaction $H^{\pi}_{\rm int}$, and (c)(e) magnetic 
interaction $H^{\psi}_{\rm int}$. (c) corresponds to $J/K=0.1$ in the ferromagnetic region and 
(e) is for $J/K=0.9$ in the antiferromagnetic region. Corresponding dominant angular-moment components 
are plotted in log-log scale in (b) and in semi-log scale in (d) and (f).}
\label{interaction}
\end{figure}

The analytical results show that the interaction consists of the part mediated by 
nonmagnetic excitation $H_{\rm int}^{\pi}$, and the part mediated by magnetic 
excitation $H_{\rm int}^{\psi}$. One remarkable difference between the two parts 
is that the latter part changes sign, when either spin is reversed, while the former 
part does not. Using this property, we can separate the nonmagnetic 
and magnetic parts of the interaction as follows:
\begin{align}
H_{\rm int}^{\pi}=&\frac{1}{2}\left\{H_{\rm int}(\bm{m}_A,\bm{m}_B)
                    +H_{\rm int}(\bm{m}_A,-\bm{m}_B)\right\} \label{f2} \\
H_{\rm int}^{\psi}=&\frac{1}{2}\left\{H_{\rm int}(\bm{m}_A,\bm{m}_B)
                    -H_{\rm int}(\bm{m}_A,-\bm{m}_B)\right\} \label{f3}
\end{align}
Further, to clarify the angular dependence of each part, we performed 
 partial-wave decomposition as 
$H_{\rm int}^{\pi(\psi)}=c_0+\sum_{m=1}^{\infty}c_m
            \cos\left(m(\theta-\theta_0^{(m)})\right)$.

Let us present the results of these analyses. First, we show the nonmagnetic 
interaction $H_{\rm int}^{\pi}$ at $J/K=0.5$ in \figref{interaction}
 (a) and (b), where dependence on separation $r_{\alpha\beta}$ and 
 radial dependences of the three large components are shown. 
We can see that the dipole component $c_2(r)$ decays as $r^{-2}$, 
   and this is the dominant component. 
It agrees with the analytical result \erefn{eq262}.
In addition, the principal axis of the angular dependence coincides with that 
of the analytical result \erefn{eq262}. Next, we turn to the magnetic 
interaction $H_{\rm int}^{\psi}$ at $J/K=0.1$ in the ferromagnetic region shown 
in \figref{interaction} (c) and (d). The monopole component $c_0(r)$ 
decays exponentially with separation $r$, and is dominant, which agrees with the 
analytical result \erefn{eq314.2}. Lastly, the magnetic interaction 
$H_{\rm int}^{\psi}$ at $J/K=0.9$ in the antiferromagnetic region is shown in 
\figref{interaction} (e) and (f). The dominant component is now dipole one $c_2(r)$, 
and the monopole component $c_0(r)$ is of the same order as dipole one 
but slightly smaller. 
This agree with the analytical result \erefn{eq298.7}, and the ratio of the 
dipole and monopole components is  $c_0/c_2=0.34$, at $r=10$, 
close to the analytical result, 0.41. Again, the principal axes agree with that 
of the analytical result.

These numerical data show that the analytical results describe the impurity-impurity 
interactions correctly on qualitative level.

\section{Discussion on Spin Freezing}
 Let us now apply the results in the previous section to discussion on 
 unusual spin freezing observed in ${\rm NiGa_2S_4}$.
First we discuss the possibility that the long range impurity-impurity interaction, 
which is mediated by nonmagnetic excitation in the bulk region,
 causes freezing of the impurity magnetic moments. Based on this 
 scenario of novel spin freezing, second we discuss the anomalies in
 experiments in ${\rm NiGa_2S_4}$ concerning spin freezing, particularly persistent 
 spin dynamics below spin freezing temperature  $T_f$ and also scaling behavior 
 of $T_f$.

Several features shown in our calculations are important to realize unusual spin freezing. 
We have shown that the anisotropy of impurity magnetic moment and 
short-range and long-range impurity-impurity interactions are caused by magnetic and nonmagnetic 
excitations in the bulk region. The anisotropic long range 
interaction mediated by nonmagnetic excitation is particularly relevant to 
the spin freezing. Let us consider the case where impurities are randomly 
located and their density is small but finite. In this case the 
spatial anisotropy of the impurity-impurity biquadratic interaction yields {\it randomness} and 
{\it frustration}, which are the origins of freezing phenomena. 
Two other points are important. First, that the biquadratic part of the impurity-impurity interactions 
are long-ranged and decays as a power law, $r^{-2}$. Second, the nematic order in the bulk induces spin anisotropy 
of impurity magnetic moments and effectively spins have only discrete degrees of freedom. 
Gandolfi {\it et al}. studied Ising spins on a hyper cubic lattice with 
 random interactions, and proved that it has a thermodynamical spin glass order at any temperature, 
 if the interaction is sufficiently long-ranged\cite{GNS}. $r^{-2}$ 
 dependence on two dimensional lattices satisfies this condition. 
Spin discreteness and long-range interactions are essential factors stabilizing spin glass 
order in their theory, and our system also shares these two points. 
Therefore it is reasonable to expect that our system has a similar thermodynamically 
stable glass order, although the effect of different discrete spin symmetries
 is not clear. It is an important 
 future issue to prove this expectation. Instead, here let us point out 
 one expected feature of spin freezing. The impurity-impurity interaction 
 is biquadratic with regard to impurity magnetic moments, and therefore 
 is invariant under the local spin inversion of each impurity magnetic moment. 
 One could alternatively say that degrees of freedom involved in the interaction 
 are headless vectors, which 
 are derived from ordinary Heisenberg spins by identifying their two tips. 
 Hence, this interaction should cause the freezing of the headless vectors, rather than original 
 magnetic dipoles. To put it simply, in this freezing state, each impurity magnetic 
 moment fluctuates so that the local expectation value of the dipole moment 
 vanishes, although it tends to be parallel or antiparallel to the local easy axis which is 
 spontaneously chosen. However, it is necessary to note that the direction of this easy axis is not 
 statically fixed. Throughout this study the coordinate system in the spin space 
 is defined relative to the ordered quadrupole moment in the bulk region as a reference 
 frame. At finite temperatures this reference frame is not static. It fluctuates 
 in the time scale of the autocorrelation time of spin quadrupole moment. 
 Therefore the easy axis of the impurity magnetic moment also fluctuates 
 with the same time scale.

Next we discuss on internal magnetic field below $T_f$, on the 
basis of the freezing mechanism just proposed. As we mentioned in \S 1, 
$\mu SR$ experiments revealed the presence of randomly distributed internal 
magnetic field, which fluctuates with a time scale of ${\rm \mu s}$, and this 
dynamics of the internal field is suppressed under magnetic field 
$H_L\geq 10 {\rm mT}$. Our scenario can explain the origin of this slow 
relaxation: that is flip of the impurity magnetic 
moments along their easy axes. Note that this flipping process without energy cost 
is a unique result of the biquadratic form of impurity-impurity interaction. 
This idea can also explain the suppression 
of the dynamics by magnetic field. The local degeneracy of the two 
directions of impurity magnetic moment, which correspond to the two tips 
of the easy axis, is lifted by external magnetic field due to Zeeman energy. 
Note that not only the impurity magnetic moment but also induced magnetic moments 
around the impurity, participate in this flipping process. 
Magnetic moments are induced  around an impurity, as we revealed in \S 4. 
The radius of this region is the magnetic correlation length, 
which is divergingly large near the phase boundary, and also 
experimentally determined as about seven times the lattice constant, 
as mentioned in \S 1. 
Therefore, 
 it is likely that relatively few impurities, for example 1\% of the bulk spins, 
 induce the quasi-static magnetic moments and their slow dynamics on most 
 sites, like observed experimentally. 
 It is also a future task 
 to investigate this possibility on more quantitative ground.

\begin{figure}[b]
\centering
\includegraphics[scale=0.44,viewport=0 0 271 281]{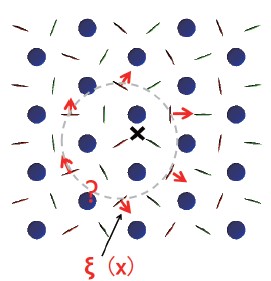}
\caption{An example of configuration of spin quadrupole moments around a quaternion vortex 
excitation in the AFQ order. On the dotted circle, we can not consistently define 
the coordinate axes, which are parallel to the director on the sublattices. }
\label{vortex}
\end{figure}
Then we present a possible explanation for the scaling behavior of the freezing 
temperature $T_f$. As mentioned in \S 1, 
$T_f$ scales with the characteristic energy scale of the low temperature 
specific heat upon controlling nonmagnetic impurity concentration. 
It implies that the freezing occurs simultaneously with some kind of transition 
which is related to the bulk order. One candidate is 
the vortex unbinding transition. Topologically stable defects exist in the AFQ 
order, as is shortly discussed in Appendix B, and therefore we expect the existence of 
a vortex binding-unbinding transition. 
This transition in the bulk affects impurity-impurity interaction. 
The AFQ order sets a reference frame of spin coordinate. For example, we defined 
$x$-axis as the director on A sublattice. In doing this, we chose one of the 
two opposite directions, both of which correspond to the same director. Such a choice is 
arbitrary, but can be done consistently in space. This situation remains 
unchanged even at finite temperatures if the system is in the vortex binding phase. 
In this phase vortices exist as bound pairs, and we can neglect their 
existence, except for the renormalization of the coupling constant, as far as 
the long range behavior of the system is concerned. The situation drastically changes, on the 
other hand, in the vortex unbinding phase, where free vortices exist. 
As shown in \figref{vortex}, we can not define the coordinate system consistently
 around a vortex. Therefore the origin of the long range impurity-impurity 
 interaction essentially breaks down in the unbinding phase. 
 Although the present study does not show explicitly yet, 
 it is natural that the interaction becomes 
short-ranged and the characteristic length scale is the order of the average 
distance between free vortices. Since it is known that two dimensional systems with short-range 
random interactions have no thermodynamical glass order\cite{AP}, we may say that 
the freezing temperature $T_f$ in our system 
coincides with the vortex unbinding transition temperature. 
This is consistent with the observed scaling behavior of $T_f$, 
since $T_f$ and the energy scale of the low-temperature specific heat are 
determined by the characteristic energy of the low-energy excitation 
in the AFQ order. We need to investigate the finite temperature problem of 
our system in order to substantiate this idea.

Now we briefly discuss the effect of the quantum fluctuation. 
One way to include this is spin-wave like approach starting with the deformed nematic order. 
We can construct a bosonic Hamiltonian in this vacuum with extended Holstein-Primakoff 
transformation, which is analogous to those adopted for the AFQ ordered system without impurities\cite{TsuneAri}. 
As stated in \S 2.2, the quantum correction to the interaction energy arises from some exchange process 
of the bosonic excitations. 
Even without quantitative investigation, we can 
predict basic characteristics of this quantum corrections. 
Reflecting the spontaneous AFQ order, there are gapless Goldstone modes with linear dispersion\cite{TsuneAri}. 
This implies the boson exchange process yields long-range (power-law decay) interactions. 
Furthermore, near the gapless point the excitations have nonmagnetic character\cite{TsuneAri}, which indicates 
the dominant long-range interaction is nonmagnetic one, while magnetic bilinear interactions would be subdominant. 
Therefore we expect the basic characteristics of the impurity-impurity interaction and also the 
discussion on spin freezing will not be changed even if we include the quantum corrections to 
the interaction.

Finally we shortly comment on other open issues. 
The models in the present study, both of the bilinear-biquadratic model for the bulk part
 and the triangular bond disorder for the impurity part, are basically 
 phenomenological. Therefore we have to verify these models 
 on a microscopic point of view. 
 
\section{Summary}
We have studied the effect of triangular-shaped ferromagnetic bond disorder 
in the $S=1$ bilinear-biquadratic model on the triangular lattice, in the parameter 
region where the antiferro quadrupolar order is realized. 
We have shown that coupling between impurity magnetic moment and magnetic and nonmagnetic 
excitations in the bulk yields several kinds of anisotropy of impurity magnetic moment, 
depending on the coupling constant of the bulk. We have also demonstrated the existence of biquadratic, 
spatially anisotropic and long-range interaction between impurity magnetic moments, and 
determined their effective coupling constants. 
This interaction is mediated by nonmagnetic excitation in the bulk. 
Based on these, we have presented the possibility of glass-like 
freezing of impurity magnetic moments due to this interaction. This scenario can explain the 
unusual spin freezing observed in ${\rm NiGa_{2}S_{4}}$ with persistent spin dynamics and the 
scaling behavior of freezing temperature with the energy scale of the bulk.

\section*{Acknowledgement}
The authors thank S. Nakatsuji, K. Ueda and A. V. Chubukov for stimulating discussions. 
This study was supported by Grant-in-Aid for Scientific Research on Priority Areas "Novel States of Matter
Induced by Frustration" (19052003). 

\appendix
\section{Extension to General Configuration of Impurities}
In \S 4 and \S 5 we have considered the case that an the impurity occupies the position $E$ in 
\figref{unit_cell}. Here we show results for 
the other positions. 
One impurity problem is explained in A.1 and impurity interactions 
are given in A.2 for general cases of pair configuration.

\subsection{One Impurity Problem}
Here we extend the calculation of the ground state field configuration around 
a single impurity in \S 4.2 and \S 4.3, 
when the impurity occupies a general position in \figref{unit_cell}. 

As for the nonmagnetic part, the ground state configuration of nonmagnetic 
excitation  $\bm{\pi}$ is given by \eref{eq114} and \erefn{eq134}, 
for the configuration $E$ in \figref{unit_cell}. 
For example, in the case of 
$C_3$, we can obtain the ground state by $\frac{2}{3}\pi$ rotation in real 
space from that in the case of $E$. 
This corresponds to the change of the lattice vectors 
as 
$\hat{\bm{\kappa}}_1\rightarrow\hat{\bm{\kappa}}_2,\ 
\hat{\bm{\kappa}}_2\rightarrow\hat{\bm{\kappa}}_3,\ 
\hat{\bm{\kappa}}_3\rightarrow \hat{\bm{\kappa}}_1$.
The other cases are also similarly obtained. 
It is natural to relate the configurations of the impurity, to the 
elements of the trigonal point group ${\rm C_{3v}}=\{E,C_3,C_3^{-1},\sigma_1,\sigma_2,\sigma_3\}$. 
That is why we introduced the notation for the configuration.

Taking advantage of this relationship, the general expression for 
the dipole moment \erefn{eq134} is given as:
\begin{equation}
\bm{\mu}_{a}^{\pi}=\mu_0^{\pi}\sum_{b,c}\frac{|\epsilon_{abc}|}{2}m_b 
                  m_c\hat{\kappa}_{a}^h, \label{eq146}
\end{equation}
where $h\in{\rm C_{3v}}$ denotes the impurity position and
\begin{equation}
\left(
\begin{array}{c}
\hat{\bm{\kappa}}_1^h \\
\hat{\bm{\kappa}}_2^h \\
\hat{\bm{\kappa}}_3^h
\end{array}
\right)
=D_h\left(
\begin{array}{c}
\hat{\bm{\kappa}}_1 \\
\hat{\bm{\kappa}}_2 \\
\hat{\bm{\kappa}}_3
\end{array}
\right). \label{eq147}
\end{equation}
Here $D_h$ is a three-dimensional representation of ${\rm C_{3v}}$:
\begin{align}
&D_E&=\left(
\begin{array}{ccc}
1&0&0 \\ 0&1&0 \\ 0&0&1
\end{array}
\right),\ 
&D_{C_3}&=\left(
\begin{array}{ccc}
0&1&0 \\ 0&0&1 \\ 1&0&0
\end{array}
\right) \nonumber \\
&D_{C_3^{-1}}&=\left(
\begin{array}{ccc}
0&0&1 \\ 1&0&0 \\ 0&1&0
\end{array}
\right),\ 
&D_{\sigma_1}&=\left(
\begin{array}{ccc}
1&0&0 \\ 0&0&1 \\ 0&1&0
\end{array}
\right) \nonumber \\
&D_{\sigma_2}&=\left(
\begin{array}{ccc}
0&0&1 \\ 0&1&0 \\ 1&0&0
\end{array}
\right),\
&D_{\sigma_3}&=\left(
\begin{array}{ccc}
0&1&0 \\ 1&0&0 \\ 0&0&1
\end{array}
\right). \label{eq148} 
\end{align}

The magnetic part can be calculated similarly and 
we show only results. Ground state configuration of magnetic field 
$\bm{\psi}$ is given by \eref{eq160}, and monopole and dipole moments are given as 
\begin{align}
q_a^{\psi}=&q_0^{\psi}m_a,\quad \bm{\mu}_{a}^{\psi}=\mu_0^{\psi}m_a
                         \hat{\bm{\kappa}}_{a}^{'h}, &(J<K/2) \label{eq190} \\
q_a^{\psi}=&0,\quad \bm{\mu}_{a}^{\psi}=\tilde{\mu}_0^{\psi}m_a
                           \hat{\bm{\kappa}}_{a}^h,&(J>K/2), \label{eq190.1}
\end{align}
where we defined
\begin{equation}
\left(
\begin{array}{c}
\hat{\bm{\kappa}}^{'h}_1 \\
\hat{\bm{\kappa}}^{'h}_2 \\
\hat{\bm{\kappa}}^{'h}_3 \\
\end{array}
\right)
=\sigma_h D_h\left(
\begin{array}{c}
\hat{\bm{\kappa}}'_1 \\
\hat{\bm{\kappa}}'_2 \\
\hat{\bm{\kappa}}'_3 \\
\end{array}
\right). \label{eq191}
\end{equation}
$D_h$ denotes the three dimensional representation of ${\rm C_{3v}}$ 
introduced by \eref{eq148}, and $\sigma_h$ denotes $A_2$ representation 
of ${\rm C_{3v}}$:$\sigma_E=\sigma_{C_3}=\sigma_{C_3^{-1}}=1,\quad
\sigma_{\sigma_1}=\sigma_{\sigma_2}=\sigma_{\sigma_3}=-1 $.

Note that the ground state energy is independent of the impurity 
position for both nonmagnetic and magnetic parts.

\subsection{Interaction between Impurities}
Using the results of the previous subsection, we can calculate the interaction 
energy for general cases of impurity pair configuration. 
The impurity $\alpha$ occupies the position $h_{\alpha}$, while the partner 
impurity $\beta$ occupies the position $h_{\beta}$ in a unit cell far away 
from the impurity $\alpha$. 
As for the nonmagnetic part, the interaction energy is 
\begin{align}
H_{\rm int}^{\pi}=&\pi J_{\pi}\frac{r_0^2(\mu_0^{\pi})^2}{r_{\alpha\beta}^2}
  {\rm Tr}\left(T^{(\alpha)}f_{h_{\alpha}h_{\beta}}T^{(\beta)}\right) \label{eq300} \\
\left(f_{h_{\alpha}h_{\beta}}\right)_{ab}=&\left\{D_{h_{\alpha}}
         \left[D_{\sigma_1}\cos(2\theta+\frac{2\pi}{3})+D_{\sigma_2}
  \cos(2\theta-\frac{2\pi}{3})  \right.\right. \nonumber\\
&\left.\left. +D_{\sigma_3}\cos 2\theta\right]D_{h_{\beta}}^T
                           \right\}_{ab}\delta_{ab}, \label{eq301}
\end{align}
where matrices $T^{(x)},\ (x=\alpha,\beta)$ are defined in 
\eref{eq263}.
We can see that principal axes of dipole-dipole like anisotropy depend on 
the position of impurity pair, but the interaction has the same 
form as before.

As for the magnetic part in the ferromagnetic region, interaction energy 
\erefn{eq288} is independent of impurity pair configuration. 
In the antiferromagnetic region, the interaction energy \erefn{eq298.6} is 
replaced by the general expression
\begin{align}
H_{\rm int}^{\psi}=&\pi J_{\psi}(\tilde{\mu}_0^{\psi})^2\bm{m}^{(\alpha)}
        \cdot\left[K_2(k_{\psi}r_{\alpha\beta})f_{h_{\alpha}h_{\beta}}
                  \right.\nonumber \\
 &\left.+K_0(k_{\psi}r_{\alpha\beta})\bm{1}\right] \bm{m}^{(\beta)}. 
                           \label{eq298.61}
\end{align}
Again, difference of impurity pair configuration is reflected only in 
the change of principal axes of dipole-dipole like anisotropy.

\section{Topological Excitation in Antiferro Quadrupolar Order}
In \S 3 we mentioned the presence of topological excitations in the AFQ order, 
and it is closely related to the mechanism of spin freezing transition presented in \S 6. Here 
we briefly summarize the properties of these topological excitations. 
Detailed general arguments on topological excitations may be found, for example, in the review by Mermin\cite{Mermin}. 
In our context, topological excitations mean defects in static configuration of the order parameter which is not 
removable by any continuous transformation of configuration\cite{TopoComment}.

Since the system of our concern is two-dimensional, relevant topological 
excitations are point defects. They can not be removed by continuous transformation 
of local order parameters inside a contour enclosing the defect. The simplest example 
for this is a vortex in ferromagnetically ordered XY spins and this is characterized by an integer winding 
number. Only defects with 
the same winding number can be continuously deformed to each other. 
In general ordered media, a simple winding number is not sufficient to label a defect, but 
one can use the homotopy theory. For order parameter space $R$, 
one defines the {\it fundamental group} $\pi_1(R)$, which is generally not Abelian, and 
a topological defect can be labeled by one of its {\it conjugacy classes}. 

\begin{table}[t]
    \begin{center}
       \caption{Class multiplication table for the quaternion group, ${\rm Q}$.}
    \label{multiplication}
    \begin{tabular}{c|ccccc}
                    & $C_0$            & $\overline{C}_0$ &       $C_x$       &      $C_y$          &       $C_z$     \\  \hline
     $C_0$          & $C_0$            & $\overline{C}_0$ &       $C_x$       &      $C_y$          &       $C_z$     \\    
    $\overline{C}_0$ & $\overline{C}_0$ & $C_0$            &       $C_x$       &      $C_y$          &       $C_z$     \\
    $C_x$ & $C_x$ & $C_x$ & $2C_0+2\overline{C}_0$ & $2C_z$ & $2C_y$ \\
    $C_y$ & $C_y$ & $C_y$ & $2C_z$ & $2C_0+2\overline{C}_0$ & $2C_x$ \\
    $C_z$ & $C_z$ & $C_z$ & $2C_y$ & $2C_x$ & $2C_0+2\overline{C}_0$
    \end{tabular}
    \end{center}
 \end{table}

Order parameter space of the AFQ state is more complicated than simple cases like 
ferromagnetic order of XY spins. This is generally defined as the coset group $G/H$, where 
$G$ is the symmetry group that keeps the original Hamiltonian 
invariant, while $H$ is its subgroup that keeps the ordered state invariant. 
The bilinear-biquadratic model \erefn{i1} has a complete spherical symmetry in 
spin space and the time reversal symmetry. Therefore $G$ is ${\rm SO(3)}\times Z_2$. The AFQ state is invariant under 
180$^{\circ}$ rotations about three mutually perpendicular axes, which form 
the dihedral group $\rm{D_2}$ of order $4$, and also the time reversal operation. Therefore the order parameter space is 
$R=({\rm SO(3)}\times Z_2)/({\rm D_2}\times Z_2)=SO(3)/D_2$ or equivalently ${\rm SU(2)}/{\rm Q}$, where ${\rm Q}$ is the {\it quaternion group}. 
It is order 8 and non-Abelian, and has a two-dimensional representation: 
\begin{equation}
{\rm Q}=\{\bm{1}, -\bm{1}, i\sigma_x, -i\sigma_x, i\sigma_y, -i\sigma_y, i\sigma_z, -i\sigma_z\}, \label{topo1}
\end{equation}
where $\sigma_a (a=x,y,z)$ are the Pauli matrices. 

It is known that for a continuous and simply connected group $G$, with a discrete subgroup $H$, the 
fundamental group of $G/H$ is $\pi_1(G/H)=H$.\cite{Mermin} In our case, $G$ is the rotation group ${\rm SU(2)}$ and 
$H$ is the quaternion group ${\rm Q}$, and therefore the fundamental group of our order parameter space is 
$\pi_1(R={\rm SU(2)}/{\rm Q})={\rm Q}$. Note that ${\rm SU(2)}$ is a simply connected group but ${\rm SO(3)}$ is not. 
Topological defects are thus classified by its five conjugacy classes as \cite{Mermin}:
\begin{align}
&C_0=\{\bm{1}\},\ \overline{C}_0=\{-\bm{1}\},\nonumber\\
&C_x=\{\pm i\sigma_x\},\ C_y=\{\pm i\sigma_y\},\ C_z=\{\pm i\sigma_z\}, \label{topo2}
\end{align}
including $C_0$ corresponding to no topological defect. 
Note that three conjugacy classes contain multiple elements of ${\rm Q}$. 
This is the result of non-Abelian nature of the group ${\rm Q}$.

\begin{figure}[h]
\centering
\includegraphics[scale=0.44,viewport=0 0 411 245]{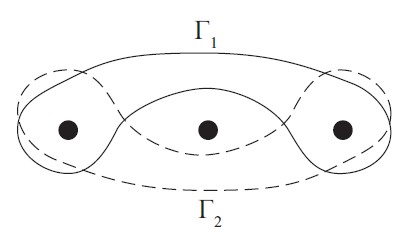}
\caption{Two contours $\Gamma_1$ and $\Gamma_2$ enclosing two defects going through the opposite sides of 
the third defect. Defects are shown by black circles. }
\label{Merge}
\end{figure}

The non-Abelian nature of the fundamental group ${\rm Q}$ implies nontrivial merging rules of topological defects. 
Consider two defects and what happens if we merge them. The point is non-uniqueness of the defect class 
for the merged defect. To define the defect class for the merged ones, 
one needs a closed path which encloses these two. The answer follows the class multiplication table for 
$\pi_1(R)={\rm Q}$ shown in \tabref{multiplication}. Note that some products contain more than one classes, reflecting 
non-commutative nature of the group ${\rm Q}$. For that case, the answer depends on the configurations 
of other defects, if present. Typical cases are shown in \figref{Merge} and the two paths both enclose the 
same two defects but they may yield different classes for the merged defect.

\ \\


\begin{thebibliography}{9}
\bibitem{TopoComment}
Topological defects may exist also in the $(d+1)$-dimensional 
effective action of a $d$-dimensional quantum Hamiltonian and might change the character of quantum mechanical excitation 
spectrum drastically. In the AFQ order, however, this type of topological defects are absent. 
This is because the space-time dimension is now 3 and one needs to examine the third homotopy group $\pi_3$, 
which is trivial for the order parameter space $SU(2)/Q$.
\end{thebibliography}
\end{document}